\documentclass[journal,10pt]{IEEEtran}
\usepackage{amsmath}
\usepackage{eqparbox}
\usepackage[usenames, dvipsnames]{color}
\usepackage{mathtools,amssymb,bm,mathabx,nccmath}
\usepackage{amstext}
\usepackage{amssymb}
\usepackage{graphicx}
\usepackage{color}
\usepackage{booktabs}
\usepackage{longtable}
\usepackage{multicol}
\usepackage{algorithmicx}
\usepackage[ruled]{algorithm}
\usepackage{algpseudocode}
\usepackage{algpascal}
\usepackage{algc}
\usepackage{amsfonts}
\usepackage{dsfont}
\usepackage{array}
\usepackage[most]{tcolorbox}
\usepackage{cases}
\usepackage{pgfplots}
\usepackage{caption}
\usepackage{subfigmat}
\usepackage{xr-hyper}
\usepackage{hyperref}
\usepackage{amsthm}
\DeclareCaptionLabelSeparator{periodspace}{.~}
\usepackage{float}
\usepackage{cite}
\usepackage{epstopdf}
\theoremstyle{remark}

\newcommand\ASTART{\bigskip\noindent\begin{minipage}[b]{0.5\linewidth}}
	
	\newcommand\AENDSKIP{\end{minipage}\bigskip}
\newcommand\AEND{\end{minipage}}
\ifCLASSOPTIONcaptionsoff
\usepackage[nomarkers]{endfloat}
\let\MYoriglatexcaption\caption
\renewcommand{\caption}[2][\relax]{\MYoriglatexcaption[#2]{#2}}
\fi
\theoremstyle{plain}
\newtheorem{thm}{\textbf{Theorem}}
\newtheorem{lem}{\textbf{Lemma}}
\newtheorem{prop}{\textbf{Proposition}}

\theoremstyle{definition}

\newtheorem{rem}{\textbf{Remark}}

\newcommand*{\rom}[1]{\expandafter\@slowromancap\romannumeral #1@}

\newcommand{\RN}[1]{%
\textup{\uppercase\expandafter{\romannumeral#1}}%
}

\usepackage{standalone}
\graphicspath{{./Figures/}}
\usepackage{times}
\usepackage[latin1]{inputenc}
\usepackage[T1]{fontenc}
\usepackage[english]{babel}
\usepackage{graphicx}
\usepackage{amsmath, amsfonts, dsfont,amssymb, graphicx, array, tabularx, booktabs,multicol}
\usepackage{amsthm}
\usepackage{epsf,epsfig}
\usepackage{setspace}
\usepackage{subfigure}

\usepackage{hyperref}
\usepackage{multirow}
\usepackage{url}
\usepackage{color}
\usepackage[nolist,printonlyused]{acronym}      
\usepackage{comment}
\usepackage{cite}
\usepackage{bm}
\usepackage{tikz}
\usetikzlibrary{decorations.pathreplacing}

\usepackage{mathtools}

\usepackage{blindtext}

\usepackage{stfloats}

\newcommand{\gf}[1]{\textcolor{black}{{#1}}}

\newtheorem{cor}{Corollary}
\newtheorem{remark}{Remark}

\newcommand{\mx}[1]{\mathbf{#1}}
\newcommand{\bs}[1]{\boldsymbol{#1}}
\providecommand{\keywords}[1]{\textbf{\textit{Index terms---}} #1}

\definecolor{amber}{rgb}{1.0, 0.49, 0.0}
\definecolor{ao}{rgb}{0.0, 0.5, 0.0}

\def\R2#1{\textcolor{black}{#1}}
\def\R3#1{\textcolor{black}{#1}}

\renewcommand{\triangleq}{\mathbin{\setstackgap{S}{0pt}\stackMath\Shortstack{\smalltriangleup\\ =}}}
\usepackage[usestackEOL]{stackengine} 

\begin{document}

\title{Sampling}
\singlespacing

\title{Unlocking the Potential of Temporal Correlations in Non-stationary Rician Aging Channels}
\title{Towards Optimal Pilot Spacing and Power Control in Multi-Antenna Systems Operating Over Non-Stationary Rician Aging Channels}
\singlespacing
\author{
 Sajad Daei$^{\dag}$, G\'{a}bor Fodor$^{\star\dag}$, Mikael Skoglund$^{\dag}$, Mikl\'{o}s Telek$^{\ddag\sharp}$\\
 \small $^\dag$KTH Royal Institute of Technology, Stockholm, Sweden. \\
\small $^\star$Ericsson Research, Stockholm, Sweden.\\

\small $^\ddag$Budapest University of Technology and Economics, Budapest, Hungary.\\
\small $^\sharp$MTA-BME Information Systems Research Group, Budapest, Hungary.
}
\maketitle
\pagestyle{plain}


\begin{acronym}
  \acro{2G}{Second Generation}
  \acro{3G}{3$^\text{rd}$~Generation}
  \acro{3GPP}{3$^\text{rd}$~Generation Partnership Project}
  \acro{4G}{4$^\text{th}$~Generation}
  \acro{5G}{5$^\text{th}$~Generation}
  \acro{AA}{Antenna Array}
  \acro{AC}{Admission Control}
  \acro{AD}{Attack-Decay}
  \acro{ADSL}{Asymmetric Digital Subscriber Line}
	\acro{AHW}{Alternate Hop-and-Wait}
  \acro{AMC}{Adaptive Modulation and Coding}
	\acro{AP}{Access Point}
  \acro{APA}{Adaptive Power Allocation}
  \acro{AR}{autoregressive}
  \acro{ARMA}{Autoregressive Moving Average}
  \acro{ATES}{Adaptive Throughput-based Efficiency-Satisfaction Trade-Off}
  \acro{AWGN}{additive white Gaussian noise}
  \acro{BB}{Branch and Bound}
  \acro{BD}{Block Diagonalization}
  \acro{BER}{bit error rate}
  \acro{BF}{Best Fit}
  \acro{BLER}{BLock Error Rate}
  \acro{BPC}{Binary power control}
  \acro{BPSK}{binary phase-shift keying}
  \acro{BPA}{Best \ac{PDPR} Algorithm}
  \acro{BRA}{Balanced Random Allocation}
  \acro{BS}{base station}
  \acro{LoS}{line of sight}
  \acro{NLoS}{non-line of sight}
  \acro{AoA}{Angle of Arrival}
  \acro{AoD}{Angle of Departure}
  \acro{CAP}{Combinatorial Allocation Problem}
  \acro{CAPEX}{Capital Expenditure}
  \acro{CBF}{Coordinated Beamforming}
  \acro{CBR}{Constant Bit Rate}
  \acro{CBS}{Class Based Scheduling}
  \acro{CC}{Congestion Control}
  \acro{CDF}{Cumulative Distribution Function}
  \acro{CDMA}{Code-Division Multiple Access}
  \acro{CL}{Closed Loop}
  \acro{CLPC}{Closed Loop Power Control}
  \acro{CNR}{Channel-to-Noise Ratio}
  \acro{CPA}{Cellular Protection Algorithm}
  \acro{CPICH}{Common Pilot Channel}
  \acro{CoMP}{Coordinated Multi-Point}
  \acro{CQI}{Channel Quality Indicator}
  \acro{CRM}{Constrained Rate Maximization}
	\acro{CRN}{Cognitive Radio Network}
  \acro{CS}{Coordinated Scheduling}
  \acro{CSI}{channel state information}
  \acro{CSIR}{channel state information at the receiver}
  \acro{CSIT}{channel state information at the transmitter}
  \acro{CUE}{cellular user equipment}
  \acro{D2D}{device-to-device}
  \acro{DCA}{Dynamic Channel Allocation}
  \acro{DE}{Differential Evolution}
  \acro{DFT}{Discrete Fourier Transform}
  \acro{DIST}{Distance}
  \acro{DL}{downlink}
  \acro{DMA}{Double Moving Average}
	\acro{DMRS}{Demodulation Reference Signal}
  \acro{D2DM}{D2D Mode}
  \acro{DMS}{D2D Mode Selection}
  \acro{DPC}{Dirty Paper Coding}
  \acro{DRA}{Dynamic Resource Assignment}
  \acro{DSA}{Dynamic Spectrum Access}
  \acro{DSM}{Delay-based Satisfaction Maximization}
  \acro{ECC}{Electronic Communications Committee}
  \acro{EFLC}{Error Feedback Based Load Control}
  \acro{EI}{Efficiency Indicator}
  \acro{eNB}{Evolved Node B}
  \acro{EPA}{Equal Power Allocation}
  \acro{EPC}{Evolved Packet Core}
  \acro{EPS}{Evolved Packet System}
  \acro{E-UTRAN}{Evolved Universal Terrestrial Radio Access Network}
  \acro{ES}{Exhaustive Search}
  \acro{FDD}{frequency division duplexing}
  \acro{FDM}{Frequency Division Multiplexing}
  \acro{FER}{Frame Erasure Rate}
  \acro{FF}{Fast Fading}
  \acro{FSB}{Fixed Switched Beamforming}
  \acro{FST}{Fixed SNR Target}
  \acro{FTP}{File Transfer Protocol}
  \acro{GA}{Genetic Algorithm}
  \acro{GBR}{Guaranteed Bit Rate}
  \acro{GLR}{Gain to Leakage Ratio}
  \acro{GOS}{Generated Orthogonal Sequence}
  \acro{GPL}{GNU General Public License}
  \acro{GRP}{Grouping}
  \acro{HARQ}{Hybrid Automatic Repeat Request}
  \acro{HMS}{Harmonic Mode Selection}
  \acro{HOL}{Head Of Line}
  \acro{HSDPA}{High-Speed Downlink Packet Access}
  \acro{HSPA}{High Speed Packet Access}
  \acro{HTTP}{HyperText Transfer Protocol}
  \acro{ICMP}{Internet Control Message Protocol}
  \acro{ICI}{Intercell Interference}
  \acro{ID}{Identification}
  \acro{IETF}{Internet Engineering Task Force}
  \acro{ILP}{Integer Linear Program}
  \acro{JRAPAP}{Joint RB Assignment and Power Allocation Problem}
  \acro{UID}{Unique Identification}
  \acro{IID}{Independent and Identically Distributed}
  \acro{IIR}{Infinite Impulse Response}
  \acro{ILP}{Integer Linear Problem}
  \acro{IMT}{International Mobile Telecommunications}
  \acro{INV}{Inverted Norm-based Grouping}
	\acro{IoT}{Internet of Things}
  \acro{IP}{Internet Protocol}
  \acro{IPv6}{Internet Protocol Version 6}
  \acro{ISD}{Inter-Site Distance}
  \acro{ISI}{Inter Symbol Interference}
  \acro{ITU}{International Telecommunication Union}
  \acro{JOAS}{Joint Opportunistic Assignment and Scheduling}
  \acro{JOS}{Joint Opportunistic Scheduling}
  \acro{JP}{Joint Processing}
	\acro{JS}{Jump-Stay}
    \acro{KF}{Kalman filter}
  \acro{KKT}{Karush-Kuhn-Tucker}
  \acro{L3}{Layer-3}
  \acro{LAC}{Link Admission Control}
  \acro{LA}{Link Adaptation}
  \acro{LC}{Load Control}
  \acro{LOS}{Line of Sight}
  \acro{LP}{Linear Programming}
  \acro{LS}{least squares}
  \acro{LTE}{Long Term Evolution}
  \acro{LTE-A}{LTE-Advanced}
  \acro{LTE-Advanced}{Long Term Evolution Advanced}
  \acro{M2M}{Machine-to-Machine}
  \acro{MAC}{Medium Access Control}
  \acro{MANET}{Mobile Ad hoc Network}
  \acro{MC}{Modular Clock}
  \acro{MCS}{Modulation and Coding Scheme}
  \acro{MDB}{Measured Delay Based}
  \acro{MDI}{Minimum D2D Interference}
  \acro{MF}{Matched Filter}
  \acro{MG}{Maximum Gain}
  \acro{MH}{Multi-Hop}
  \acro{MIMO}{multiple input multiple output}
  \acro{MINLP}{Mixed Integer Nonlinear Programming}
  \acro{MIP}{Mixed Integer Programming}
  \acro{MISO}{Multiple Input Single Output}
  \acro{ML}{maximum likelihood}
  \acro{MLWDF}{Modified Largest Weighted Delay First}
  \acro{MME}{Mobility Management Entity}
  \acro{MMSE}{minimum mean squared error}
  \acro{LMMSE}{linear MMSE}
  \acro{MOS}{Mean Opinion Score}
  \acro{ASE}{averaged spectral efficiency}
  \acro{MPF}{Multicarrier Proportional Fair}
  \acro{MRA}{Maximum Rate Allocation}
  \acro{MR}{Maximum Rate}
  \acro{MRC}{maximum ratio combining}
  \acro{MRT}{Maximum Ratio Transmission}
  \acro{MRUS}{Maximum Rate with User Satisfaction}
  \acro{MS}{mobile station}
  \acro{MSE}{mean squared error}
  \acro{MSI}{Multi-Stream Interference}
  \acro{MTC}{Machine-Type Communication}
  \acro{MTSI}{Multimedia Telephony Services over IMS}
  \acro{MTSM}{Modified Throughput-based Satisfaction Maximization}
  \acro{MU-MIMO}{multiuser multiple input multiple output}
  \acro{MU}{multi-user}
  \acro{NAS}{Non-Access Stratum}
  \acro{NB}{Node B}
  \acro{NE}{Nash equilibrium}
  \acro{NCL}{Neighbor Cell List}
  \acro{NLP}{Nonlinear Programming}
  \acro{NLOS}{Non-Line of Sight}
  \acro{NMSE}{Normalized Mean Square Error}
  \acro{NORM}{Normalized Projection-based Grouping}
  \acro{NP}{non-polynomial time}
  \acro{NRT}{Non-Real Time}
  \acro{NSPS}{National Security and Public Safety Services}
  \acro{O2I}{Outdoor to Indoor}
  \acro{OFDMA}{orthogonal frequency division multiple access}
  \acro{OFDM}{orthogonal frequency division multiplexing}
  \acro{OFPC}{Open Loop with Fractional Path Loss Compensation}
	\acro{O2I}{Outdoor-to-Indoor}
  \acro{OL}{Open Loop}
  \acro{OLPC}{Open-Loop Power Control}
  \acro{OL-PC}{Open-Loop Power Control}
  \acro{OPEX}{Operational Expenditure}
  \acro{ORB}{Orthogonal Random Beamforming}
  \acro{JO-PF}{Joint Opportunistic Proportional Fair}
  \acro{OSI}{Open Systems Interconnection}
  \acro{PAIR}{D2D Pair Gain-based Grouping}
  \acro{PAPR}{Peak-to-Average Power Ratio}
  \acro{P2P}{Peer-to-Peer}
  \acro{PC}{Power Control}
  \acro{PCI}{Physical Cell ID}
  \acro{PDF}{Probability Density Function}
  \acro{PDPR}{pilot-to-data power ratio}
  \acro{PER}{Packet Error Rate}
  \acro{PF}{Proportional Fair}
  \acro{P-GW}{Packet Data Network Gateway}
  \acro{PL}{Pathloss}
  \acro{PPR}{pilot power ratio}
  \acro{PRB}{physical resource block}
  \acro{PROJ}{Projection-based Grouping}
  \acro{ProSe}{Proximity Services}
  \acro{PS}{Packet Scheduling}
  \acro{PSAM}{pilot symbol assisted modulation}
  \acro{PSO}{Particle Swarm Optimization}
  \acro{PZF}{Projected Zero-Forcing}
  \acro{QAM}{Quadrature Amplitude Modulation}
  \acro{QoS}{Quality of Service}
  \acro{QPSK}{Quadri-Phase Shift Keying}
  \acro{RAISES}{Reallocation-based Assignment for Improved Spectral Efficiency and Satisfaction}
  \acro{RAN}{Radio Access Network}
  \acro{RA}{Resource Allocation}
  \acro{RAT}{Radio Access Technology}
  \acro{RATE}{Rate-based}
  \acro{RB}{resource block}
  \acro{RBG}{Resource Block Group}
  \acro{REF}{Reference Grouping}
  \acro{RLC}{Radio Link Control}
  \acro{RM}{Rate Maximization}
  \acro{RNC}{Radio Network Controller}
  \acro{RND}{Random Grouping}
  \acro{RRA}{Radio Resource Allocation}
  \acro{RRM}{Radio Resource Management}
  \acro{RSCP}{Received Signal Code Power}
  \acro{RSRP}{Reference Signal Receive Power}
  \acro{RSRQ}{Reference Signal Receive Quality}
  \acro{RR}{Round Robin}
  \acro{RRC}{Radio Resource Control}
  \acro{RSSI}{Received Signal Strength Indicator}
  \acro{RT}{Real Time}
  \acro{RU}{Resource Unit}
  \acro{RUNE}{RUdimentary Network Emulator}
  \acro{RV}{Random Variable}
  \acro{SAC}{Session Admission Control}
  \acro{SCM}{Spatial Channel Model}
  \acro{SC-FDMA}{Single Carrier - Frequency Division Multiple Access}
  \acro{SD}{Soft Dropping}
  \acro{S-D}{Source-Destination}
  \acro{SDPC}{Soft Dropping Power Control}
  \acro{SDMA}{Space-Division Multiple Access}
  \acro{SE}{spectral efficiency}
  \acro{SER}{Symbol Error Rate}
  \acro{SES}{Simple Exponential Smoothing}
  \acro{S-GW}{Serving Gateway}
  \acro{SINR}{signal-to-interference-plus-noise ratio}
  \acro{SI}{Satisfaction Indicator}
  \acro{SIP}{Session Initiation Protocol}
  \acro{SISO}{single input single output}
  \acro{SIMO}{Single Input Multiple Output}
  \acro{SIR}{signal-to-interference ratio}
  \acro{SLNR}{Signal-to-Leakage-plus-Noise Ratio}
  \acro{SMA}{Simple Moving Average}
  \acro{SNR}{signal-to-noise ratio}
  \acro{SORA}{Satisfaction Oriented Resource Allocation}
  \acro{SORA-NRT}{Satisfaction-Oriented Resource Allocation for Non-Real Time Services}
  \acro{SORA-RT}{Satisfaction-Oriented Resource Allocation for Real Time Services}
  \acro{SPF}{Single-Carrier Proportional Fair}
  \acro{SRA}{Sequential Removal Algorithm}
  \acro{SRS}{Sounding Reference Signal}
  \acro{SU-MIMO}{single-user multiple input multiple output}
  \acro{SU}{Single-User}
  \acro{SVD}{Singular Value Decomposition}
  \acro{TCP}{Transmission Control Protocol}
  \acro{TDD}{time division duplexing}
  \acro{TDMA}{Time Division Multiple Access}
  \acro{TETRA}{Terrestrial Trunked Radio}
  \acro{TP}{Transmit Power}
  \acro{TPC}{Transmit Power Control}
  \acro{TTI}{Transmission Time Interval}
  \acro{TTR}{Time-To-Rendezvous}
  \acro{TSM}{Throughput-based Satisfaction Maximization}
  \acro{TU}{Typical Urban}
  \acro{UE}{User Equipment}
  \acro{ULA}{Uniform Linear Array}
  \acro{UEPS}{Urgency and Efficiency-based Packet Scheduling}
  \acro{UL}{uplink}
  \acro{UMTS}{Universal Mobile Telecommunications System}
  \acro{URI}{Uniform Resource Identifier}
  \acro{URM}{Unconstrained Rate Maximization}
  \acro{UT}{user terminal}
  \acro{VR}{Virtual Resource}
  \acro{VoIP}{Voice over IP}
  \acro{WAN}{Wireless Access Network}
  \acro{WCDMA}{Wideband Code Division Multiple Access}
  \acro{WF}{Water-filling}
  \acro{WiMAX}{Worldwide Interoperability for Microwave Access}
  \acro{WINNER}{Wireless World Initiative New Radio}
  \acro{WLAN}{Wireless Local Area Network}
  \acro{WMPF}{Weighted Multicarrier Proportional Fair}
  \acro{WPF}{Weighted Proportional Fair}
  \acro{WSN}{Wireless Sensor Network}
  \acro{WWW}{World Wide Web}
  \acro{XIXO}{(Single or Multiple) Input (Single or Multiple) Output}
  \acro{ZF}{zero-forcing}
  \acro{ZMCSCG}{Zero Mean Circularly Symmetric Complex Gaussian}
\end{acronym}

\begin{abstract}
Several previous works have addressed the inherent trade-off between allocating resources in the power and time domains to pilot and data signals in multiple input multiple output systems over block-fading channels. In particular, when the channel changes rapidly in time, channel aging degrades the performance in terms of spectral efficiency without proper pilot spacing and power control. Despite recognizing non-stationary stochastic processes as more accurate models for time-varying wireless channels, the problem of pilot spacing and power control in multi-antenna systems operating over non-stationary channels is not addressed in the literature. In this paper, we address this gap by introducing a refined first-order autoregressive model that exploits the inherent temporal correlations over non-stationary Rician aging channels. We design a multi-frame structure for data transmission that better reflects the non-stationary fading environment than previously developed single-frame structures. Subsequently, to determine optimal pilot spacing and power control within this multi-frame structure, we develop an optimization framework and an efficient algorithm based on maximizing a deterministic equivalent expression for the spectral efficiency, demonstrating its generality by encompassing previous channel aging results. Our numerical results indicate the efficacy of the proposed method in terms of spectral efficiency gains over the single frame structure.

\end{abstract}  

\section{Introduction}
As several previous works have pointed out, the performance of the uplink of \ac{MIMO} systems
depends critically on the quality of the available \ac{CSIR}
\cite{Medard:00, Hassibi:03, Vikalo:04}.
\gf{Therefore}, assuming a block fading reciprocal channel, a finite
number of symbols in the time and frequency domains are typically made available for \ac{CSIR} acquisition,
while the remaining symbols are used for data transmission \cite{Marzetta:06}, \cite{Marzetta:13}.
Also, under a fixed power budget, pilot symbols reduce the transmitted
energy for data symbols, as it has been shown in \cite{KA:05}, \cite{kim2006cth03}, where
the near-optimal \ac{PDPR} for various pilot patterns and receiver structures have been derived.
Subsequently, reference \cite{Jindal:10} has optimized the pilot overhead
for single-user wireless fading channels, and studied how various system parameters
of interest (e.g. fading rate, \ac{SNR}) depend on this overhead. Reference \cite{Fodor:16}
has derived analytical expressions for the achievable \ac{SE}
as a function of the \ac{PDPR} in the presence correlated channels. 
Note the above papers have considered block-fading channels, where the subsequent channel realizations
are assumed to be independent and drawn from the same distribution.
\gf{For block-fading channel models, important prior works have developed the conceptually elegant and practically useful concept of the deterministic equivalent \ac{SINR}  that
helps to derive analytical expressions for the achievable \ac{SINR} and spectral efficiency of multi-antenna systems \cite{Hoydis:13,Mallik:18,Hanlen:2012,Couillet:2012, Wen:2013}} . 

In contrast, some previous works have suggested that in the
case when the subsequent channel realizations are correlated, the temporal correlation structure can be
exploited to improve the quality of the channel estimates \cite{Fodor:2021, Fodor:22}.

Another set of important related works have considered the problem of acquiring \ac{CSIR} in the
presence of channel aging \cite{Abeida:10, Hijazi:10, Truong:13, Kong:2015, Yuan:20, Kim:20, Fodor:23}.
Channel aging refers to the evolution of the channel between subsequent estimation instances, and
can be conveniently modelled as a first-order or higher-order stationary \ac{AR} process, whose states can
be estimated and predicted using Kalman or Wiener filters \cite{Truong:13, Kim:20}. 

\gf{Specifically, the work in \cite{Truong:13} explored the effects of channel aging on the \ac{SINR} performance in 
single input multiple output
uplink systems and 
multiple input single output
downlink systems, and derived closed-form expressions for the deterministic equivalent \ac{SINR} under channel aging. 
Extending that work, reference \cite{Fodor:23} considered a more general \ac{AR} model, 
assuming a specific state transition matrix in the form of exponential decaying matrices to formulate a closed-form expression for the deterministic 
equivalent \ac{SINR} in \ac{AR} aging channels.}

As it has
been shown in, for example, \cite{Savazzi:09, Savazzi:09B, Fodor:23}, channel aging gives rise
to the inherent problem of pilot spacing or frame size dimensioning. Frame size dimensioning refers
to determining how densely pilot symbols should be inserted in the flow of data symbols such that
the available \ac{CSIR} is properly updated at the expense of some pilot overhead. A key insight
provided by these papers is that the performance of the uplink of \ac{MIMO} systems depends both
on the \ac{PDPR} and the frame size, and both should be properly set such that a good balance
between \ac{CSIR} acquisition and communication is achieved in the presence of some power budget
and channel aging.

Along a related research line, several recent works have shown that in many wireless communication
scenarios, the evolution of the wireless channel is advantageously modeled as a non-stationary
stochastic process, whose first and second order statistics -- including the mean and covariance matrices of the channels --
evolve in time \cite{non_stationary_model, Banerjee:20, Iimori:21, Loschenbrand:22, Song:22,cheng2022channel, zou2023joint, bian2021general}.
Specifically, the measurement results obtained in a \ac{MIMO} system and reported in \cite{Loschenbrand:22,careem2020real}
suggest that the propagation conditions and thereby the wireless channels are non-stationary in space and
time even under low mobility due to varying mobile user positions and changes in the propagation environment.
Non-stationary channel aging is also considered in \cite{Iimori:21}, where a convergent \ac{MMSE} beamforming
algorithm is proposed, which outperforms state-of-the-art beamforming methods in the downlink of
\ac{MIMO} systems. However, these papers do not treat the inherent trade-offs discussed above.


In the light of the above results on the \ac{PDPR}, frame size dimensioning (pilot spacing) and the
importance of modelling the evolution of the wireless channels as non-stationary processes, in this
paper we argue that \ac{CSIR} acquisition should be designed for \ac{MIMO} systems operating
over non-stationary wireless systems. Specifically, in this paper, we consider the uplink of
a \ac{MU-MIMO} system operating over non-stationary aging Rician channels. 
In these channels, both the mean and covariance matrices of the channel age in time, 
which can lead to a degradation in performance in terms of \ac{SE}. 

Therefore, in this paper, we first introduce a refined first-order \ac{AR} model, meticulously tailored to capture the intricacies of time-varying Rician non-stationary aging channels. 
\gf{This model harnesses the underlying temporal correlations of the channel by allowing the channel covariance matrices to change in time.}

Next, we introduce a multi-frame structure, where each frame consists of one pilot time slot and the remaining slots are allocated for data transmission, as illustrated in Figure \ref{fig:framework}. 
The frame sizes (pilot spacing), the number of frames, and \ac{PDPR} are not predetermined. 
\gf{We propose} an analytical optimization framework to maximize 
\gf{the deterministic equivalent \ac{SE}} under specific power constraints.
This framework identifies critical parameters, such as frame size, the number of frames, and the optimal power allocation between data and pilot \gf{symbols}. 
Significantly, this optimization relies solely on \gf{the temporal dynamics of the channel}, 
without dependence on measurements or channel and data estimates. 
Consequently, \gf{the proposed design} can be executed at the transmitter side. 
Central to our work is the pivotal question of determining the optimal rate for updating 
\gf{the \ac{CSIR}}, that is the optimal pilot spacing.
Our proposed optimization framework explicitly addresses this fundamental question.

Furthermore, the outcomes of extensive numerical experiments validate the efficacy of our methodology. They underscore its profound impact on optimizing pilot power, data power, frame sizes, and the number of frames, 
which are all critical elements in the quest for enhanced \ac{SE}. 

\subsection{Contributions and Key Differences Compared with Prior Works}
In this subsection, we highlight the main contributions of this paper and discuss the key differences 
compared with prior works.
\begin{enumerate}
     \item \textbf{\gf{Refined} \ac{AR} Channel Modeling}: We propose a \gf{refined} representation of \gf{the time-varying channel by a refined \ac{AR} channel model} in Proposition \ref{prop.channel_model}, which not only exploits temporal channel correlations with previous time instances, but it also takes into account the innovative channel information of the current time. The state transition matrix depends explicitly on the so-called correlation matrix, which is equivalent to the auto-correlation function of the normalized centered channel.
     
     The proposed \ac{AR}  model covers the Rician non-stationary aging channel, which is more general than the Rayleigh stationary case provided in \cite{Fodor:23,Truong:13,Savazzi:09,Savazzi:09B}. Moreover, in the special case of Rayleigh stationary fading investigated in \cite{Fodor:23,Savazzi:09, Savazzi:09B}, it is not obvious what is the optimal \ac{AR} state transition matrix and what is the optimal covariance matrix of the error of \ac{AR} modeling in general. In contrast, our model establishes explicit links between the \ac{AR} state transition matrix and the so-called correlation matrix that is accessible in practice.

     \item \textbf{Deterministic Equivalent Spectral Efficiency}: We propose an analytical framework in Theorems \ref{thm.Instantanous_SINR} and \ref{thm.stiel} for Rician aging non-stationary fast-fading channels, utilizing concentration inequalities derived from random matrix theory tools. This framework provides an exact deterministic approximation of the average achievable \ac{SE}, termed the deterministic equivalent \ac{SE}, as the number of \ac{BS} antennas grows significantly. The outcome is expressed as a function of various parameters, including frame sizes, the number of frames, pilot and data powers, the number of receiver antennas, and the Rician K factor.

    In the context of stationary Rayleigh block fading channels, prior works such as \cite{Hoydis:13} and \cite{Wagner:12} obtained approximations for the deterministic equivalent \ac{SINR}, while \cite{Fodor:23,Truong:13} established an upper bound for the average achievable \ac{SE} in stationary Rayleigh aging channels. Notably, in our \ac{SE} derivations, we employ the \ac{MMSE} estimator for data symbol estimation and incorporate an aging-aware \ac{MMSE} receiver combiner. This approach contrasts with \cite{Fodor:23}, which relies on least square estimation for the optimal receiver combiner. The distinction underscores that our method effectively utilizes the pre-existing aging information of the channels to determine the optimal receiver combiner for data estimation, providing a crucial advantage over the approach in \cite{Fodor:23}.
     
    \item \textbf{Optimal Frame Design and Power Allocation in Multi-frame Data Transmission}: Our approach provided in Section \ref{Sec:Proposed_scheme} introduces a multi-frame framework, as illustrated in Figure \ref{fig:framework}, wherein each frame comprises one pilot time slot and the remaining slots are allocated for data transmission. This flexible framework accommodates an arbitrary number of frames and data time slots within each frame. Subsequently, we formulate an optimization problem with the objective of maximizing the deterministic equivalent \ac{SE} averaged across all data time slots and frames, while satisfying some power constraints. This optimization task entails determining optimal values for various parameters, including the number of frames, frame sizes, as well as pilot and data powers. The proposed problem inherently falls under mixed-integer programming, a category generally recognized as NP-hard. To address this complexity, we provide a novel, efficient algorithm in Algorithm \ref{alg.optweights} named OptResource designed to tackle the intricacies of solving the optimization problem.

While \cite{Fodor:23} exclusively optimizes pilot spacing (frame size) in Rayleigh stationary fading scenarios, making the implicit assumption of equal frame sizes, predetermined knowledge of the number of frames (number of pilots), pilot and data powers, our method goes beyond by optimizing the number of frames, diverse frame sizes, and the allocation of pilot and power in the broader and more complex context of Rician aging non-stationary environments. Notably, even in stationary environments, our proposed multi-frame optimization outperforms the single-frame optimization presented in \cite{Fodor:23} in terms of \ac{SE}. Furthermore, it is crucial to highlight that \cite{Fodor:23} seeks to maximize an upper-bound on the average \ac{SE} when determining optimal pilot spacing, while our proposed framework exactly maximizes the average achievable \ac{SE}. This distinction is significant, as maximizing upper-bounds does not guarantee the maximization of average \ac{SE}.

\item    \textbf{Optimization at the Transmitter or Receiver}: Our findings provided in Section \ref{Sec:Proposed_scheme} and \ref{Sec:simulations} indicate that interference components such as path loss, pilot and data powers, and Doppler frequencies impact the achieved deterministic spectral efficiency but are irrelevant in determining optimal frame design. This suggests that all optimization tasks regarding frame design can be carried out at the transmitter side  while optimal power-domain resources can be calculated at the receiver and the receiver can inform the transmitter about them by some control or feedback loops. To the best of our knowledge, prior works such as \cite{Fodor:23, Savazzi:09, Savazzi:09B} assert the achievement of optimal pilot spacing at the receiver side.
    
    \item \textbf{Accessible Correlation Information in Non-stationary Environments}:
    We derive explicit formulas in Proposition \ref{prop.rician} for the required correlation matrix regarding the proposed framework in Section \ref{Sec:Proposed_scheme}, and show that they are accessible beforehand in practical non-stationary Rician time-varying environments. Notably, in the special scenario of stationary Rayleigh aging channels, our findings align with the well-known Jakes model, as previously derived in \cite{baddour2004accurate}.
\end{enumerate}

\subsection{Outline}
The paper is organized as follows. Section \ref{Sec:Model} 
proposes a channel model that exploits temporal correlations, and presents the measurement model at the receiver. Section \ref{Sec:Proposed_scheme} specifies the road to obtain the instantaneous (random) \ac{SINR} and \ac{SE} and proposes a deterministic expression for the achievable \ac{SE} based on random matrix theory tools that concentrates around the random instantaneous \ac{SE}. In Section \ref{Sec:proposed_strategy}, we propose an optimization problem that takes into account this deterministic equivalent \ac{SE} and finds the optimal values of frame sizes, number of frames and pilot and data powers. We also propose a heuristic algorithm to find the optimal values of frame size, number of frames and pilot and data powers. The paper is concluded in Section \ref{sec:conclusion}.

\subsection{Notation}
Scalars are denoted by italic letters, while matrices and vectors are shown by bold upright letters.
$\mx{I}_N$ denotes the identity matrix of size $N$. $\textbf{vec}$ stands for the column stacking vector operator that transforms a matrix $\mx{X}\in\mathbb{C}^{M\times N}$ into its vectorized version $\mx{x}\triangleq \textbf{vec}(\mx{X})\in\mathbb{C}^{M N\times 1}$. 
For a random process $\mx{h}(t)\in\mathbb{C}^{N_r\times 1}$ that shows the channel at the time slot $t\in \mathbb{N}^{+}$, 
$\overline{\mx{h}}(t)\triangleq \mathds{E}[\mx{h}(t)]$ denotes the channel mean and $\widetilde{\mx{{h}}}(t)\triangleq\mx{h}(t) -\overline{\mx{{h}}}(t)$ denotes the centered channel.
Furthermore, the autocovariance and autocorrelation are given by $\mx{C}_{\mx{h}}(t)\triangleq \mathds{E}\left[ \widetilde{\mx{{h}}}(t) \widetilde{\mx{{h}}}(t)^{\rm H}\right]$ and $\mx{R}_{\mx{h}}(t)\triangleq \mathds{E}\left[ \mx{h}(t) \mx{h}(t)^{\rm H}\right]$. The cross-covariance of the channel between times $t_1$ and $t_2$ is shown by $\mx{C}_{\mx{h}}(t_1,t_2)\triangleq\mathds{E}\left[\widetilde{\mx{h}}(t_1)\widetilde{\mx{h}}^{\rm H}(t_2)\right]$.
The normalized centered channel is shown by $\mx{h}^{\prime}(t)\triangleq \mx{C}^{-\tfrac{1}{2}}_{\mx{h}}(t)\widetilde{\mx{h}}(t)$. 
To avoid complexity, we occasionally eliminate the time dependency when referring to autocovariance or autocorrelation matrices. $\mx{e}_k\in \mathbb{R}^{N\times 1}$ refers to the $k$-th unit vector, with components equal to zero except for the $k$-th component that is one. $\mx{1}_N\in\mathbb{R}^{N\times 1}$ is an all-one vector of size $N$. $j\triangleq\sqrt{-1}$ is the complex imaginary unit. $\lambda_{\max}(\mx{A})$ stands for the spectral radius of matrix $\mx{A}$. $x_l$ stands for the $l$-th element of $\mx{x}\in \mathbb{C}^{N\times 1}$.
$[\mx{A}]_{k,l}$ denotes the $(k,l)$-th element of the matrix $\mx{A}$. $J_0(\cdot)$ is the Bessel function of zero kind and is defined as $J_0(x)\triangleq \tfrac{1}{2\pi}\int_{-\pi}^{\pi} {\rm e}^{-j x \sin(t)}{\rm d}t$. The inner product between two matrices $\mx{A}$ and $\mx{B}$ is defined as $\langle \mx{A}, \mx{B}\rangle\triangleq {\rm trace}(\mx{A}^{\rm H} \mx{B})$. For a vector $\mx{x}\in \mathbb{C}^{N\times 1}$, the $\ell_2$ norm is defined as $\|\mx{x}\|_2\triangleq\sqrt{\sum_{l=1}^N |x_l|^2}$.  The Kronecker product between two vectors $\mx{x}\in\mathbb{C}^{N\times 1}$ and $\mx{y}\in\mathbb{C}^{M\times 1}$ is shown by $\mx{x}\otimes \mx{y} \in\mathbb{C}^{MN\times 1}$. $\mathds{E}_{x,y}$ denotes the expectation operator over the joint distribution of random variables $x$ and $y$. To simplify notation, throughout the paper, we tag User-1 as the intended user, and will sometimes drop index $k=1$ when referring to the tagged user.

\section{System Model}
\label{Sec:Model}
\subsection{Channel Model}
Consider a time $t\in\mathbb{N}^{+}$ represented as an integer multiple of the symbol duration, denoted by $T$. Without loss of generality, we assume that $T=1$ throughout the paper. The wireless communication link between a user with a single antenna and a \ac{BS} equipped with $N_r$ antennas at time slot $t$, represented by $\mx{h}(t)\in\mathbb{C}^{N_r\times 1}$, is typically characterized as a non-stationary complex stochastic process. Specifically, the mean $\overline{\mx{h}}(t)$, autocovariance $\mx{C}_{\mx{h}}(t)$ and cross-covariance of $\mx{h}(t)\in\mathbb{C}^{N_r\times 1}$ between arbitrary time slots $t_1$ and $t_2$ denoted by $\mx{C}_{\mx{h}}(t_1,t_2)$ change with time.

 In fact, the cross-covariance function depends not only on the time lag (i.e., $t_2-t_1$) but also on the reference time $t_1$. In the following proposition, we state how the time-varying channel at time $t+1$ can be modeled as an \ac{AR} process exploiting the temporal correlations with previous time $t$.

\begin{prop}
\label{prop.channel_model}
Let $t$ be any time slot and $\mx{h}(t)\in\mathbb{C}^{N_r\times1}$ be a random channel at time $t$.
The channel at time $t+1$ can be represented
by the channel at time $t$ as follows:

\begin{align}
\label{eq:AR1}
&\widetilde{\mx{h}}(t+1)
&=\underbrace{ \mx{A}(t)\widetilde{\mx{h}}(t) }_{\textup{Correlation information}}+ \underbrace{\bs{\xi}(t+1)}_{\textup{Innovative information}},    
\end{align}
where the state transition matrix $\mx{A}(t)$ is defined as
\begin{align}
 \mx{A}(t) \triangleq  \mx{C}_{\mx{h}}(t+1,t)\mx{C}_{\mx{h}}^{-1}(t), 
\end{align}
and the \ac{AR} noise
$\bs{\xi}(t+1)$ is distributed as $\mathcal{CN}(\mx{0},\mx{\Theta}(t+1))$ with
\begin{align}
   \mx{\Theta}(t+1)= \mx{C}_{\mx{h}}(t+1)-\mx{C}_{\mx{h}}(t+1,t) \mx{C}^{-1}_{\mx{h}}(t)\mx{C}_{\mx{h}}(t,t+1).
\end{align}
\end{prop}

\begin{rem}(Channel modeling intuition)
The channel at time $t+1$ in \eqref{eq:AR1} consists of two terms, which are independent from each other. The first term specifies the temporal correlation of the channel at time $t+1$ with
that of time $t$, while the second term provides the innovative channel information of the current time. However, the covariance of the second term is related to the temporal correlation, as 
it specifies the error caused by representing the channel vector at time $t+1$, based on the channel at time $t$. For a given integer $n$, the \ac{AR} model in \eqref{eq:AR1} implies that the spectral radius of the correlation matrix between $t+n$ and $t$, that is the cross-covariance of the normalized centered channel, and is denoted by

{\footnotesize{
\begin{align}\label{eq:correlation_mat}
&\mx{P}_{\mx{h}}(t+n,t)\triangleq \mathds{E}[ {\mx{h}}^{\prime}(t+n) {{\mx{h}}^{\prime}}^{\rm H}(t)] = \mx{C}_\mx{{h}}^{-\tfrac{1}{2}}(t+n)\mx{C}_{\mx{h}}(t+n,t)\mx{C}_{\mx{h}}^{-\tfrac{1}{2}}(t),
\end{align}}}
must be less than one. This makes sure that the correlation between the channel values decreases as they become further apart in time. This makes sense, because as time progresses, environmental conditions, user mobility, and other factors introduce variations in the channel, leading to decreasing correlation between channel observations made at different time instances.
\end{rem}

\begin{figure}
    \centering
    \includegraphics[scale=.2]{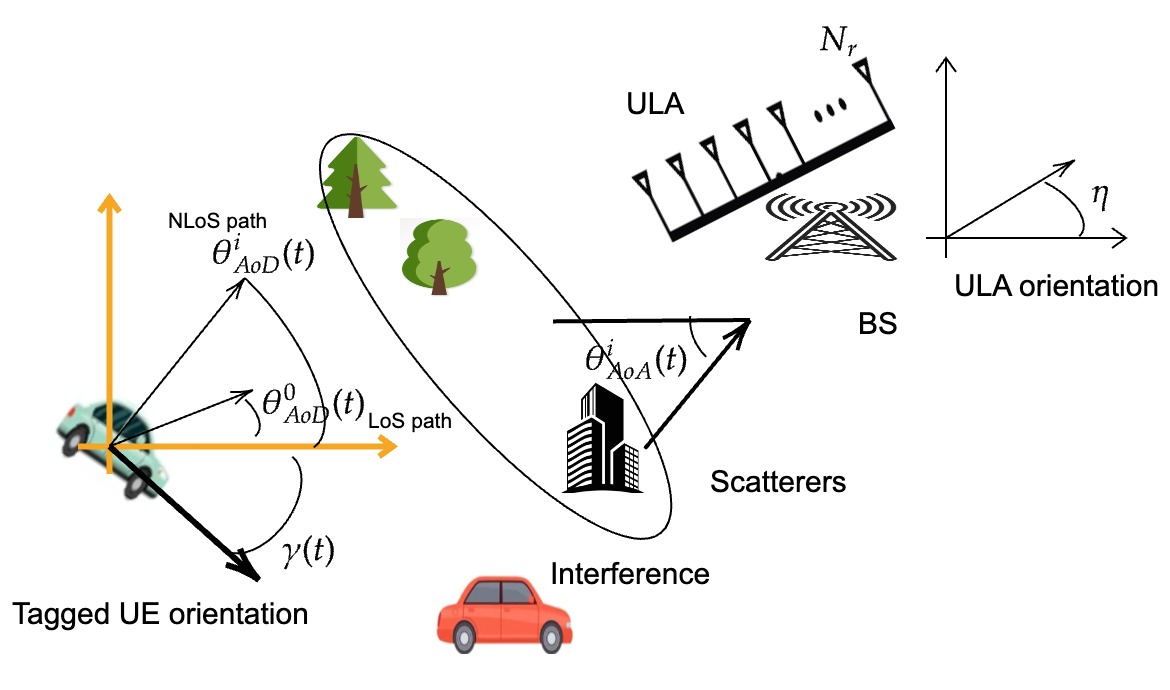}
    \caption{A typical example of Rician time-varying channel in vehicular networks.}
    \label{fig:rician_model}
\end{figure}

\subsection{Correlation \gf{M}atrix in \gf{T}ime-\gf{V}arying \gf{E}nvironments}


The correlation matrix of the channel is typically influenced by several factors, including the propagation geometry, user's velocity, and antenna characteristics. 
Consider a tagged vehicle moving towards direction $\gamma(t)$ at time $t$  
with the speed $\nu(t)$ as shown in Figure \ref{fig:rician_model}. Suppose that the \ac{BS} is equipped with an \ac{ULA} of $N_r$ number of antennas that is aligned in direction $\eta$ (see Figure \ref{fig:rician_model}). The distance between antenna elements is denoted by $d$. The environment between the tagged \ac{UE} and the \ac{BS} is composed of $s(t)$ number of scatterers and the Doppler frequencies of the \ac{LoS} and \ac{NLoS} paths are obtained by
\begin{align}
     &f_d^0(t)=\tfrac{\nu(t)\cos(\gamma(t)-\theta_{\rm AoD}^{0}(t))}{\lambda}, f_d^i(t)=\tfrac{\nu(t)\cos(\gamma(t)-\theta_{\rm AoD}^{i}(t))}{\lambda},
\end{align}
where $i=1,..., s(t)$, $\lambda\triangleq \tfrac{c}{f_c}$ is the wavelength of the source, $c=3\times 10^8$ m/s, $f_c$ is the carrier frequency, $\theta_{\rm AoD}^{0}(t))$ is the \ac{AoD} for the \ac{LoS} angle and $\theta_{\rm AoD}^{i}(t)), i=1,..., s(t)$ are the \acp{AoD} for \ac{NLoS} paths. For this case, we state the following proposition, which will be useful in the sequel.
\begin{prop}\label{prop.rician}
Assume that the number of scatterers is sufficiently large ($s(t)\rightarrow \infty$) and the \ac{AoD} and \acp{AoA} are distributed according to some known distributions. Then, the $(k,l)$ element of the correlation matrix $\mx{P}_{\mx{h}}(t_1,t_2)$ can be obtained as
\begin{align}
&\rho_{k,l}(t_1,t_2) \triangleq [\mx{P}_{\mx{h}}(t_1,t_2)]_{(k,l)}=\mathds{E}_{\theta_{\rm AoD}(t_1)),\theta_{\rm AoD}(t_2))}\Big[\nonumber\\
    &
   {\rm e}^{j\tfrac{2\pi}{\lambda}\big[t_1\nu(t_1)\cos(\gamma(t_1)-\theta_{\rm AoD}(t_1))-t_2\nu(t_2)\cos(\gamma(t_2)-\theta_{\rm AoD}(t_2))\big]}\Big]
   \nonumber\\
   &\mathds{E}_{\theta_{\rm AoA}(t_1)),\theta_{\rm AoA}(t_2))}\Big[{\rm e}^{j\tfrac{2\pi d}{\lambda}\big[k\cos(\eta-\theta_{\rm AoA}(t_1))-l\cos(\eta-\theta_{\rm AoA}(t_2))\big]}
   \Big]\nonumber\\
   &\triangleq\rho_{\rm temporal}(t_1,t_2)\rho_{\rm spatial}(t_1,t_2,k,l).
\end{align}
\end{prop}
Proof. See Appendix \ref{proof.rician}.

This model aligns with the widely recognized U-shaped band-limited power spectral density, resulting from the assumption that the propagation environment is two-dimensional (2-D) with non-isotropic scattering (more detailed can be found in \cite{non_stationary_model, baddour2004accurate, jakes1974mobile}).
If the distributions of \acp{AoA} and \ac{AoD} are known in advance, then the correlation $\rho_{k,l}(t_1,t_2)$ can be fully calculated by numerical integration. It is crucial to emphasize that having a predictive understanding of the evolving behavior of the channel covariance over time, as explored in studies such as \cite{careem2020real,Iimori:21,matz2005non,paier2008non}, can significantly benefit our approach. Utilizing the direct relationship provided by Equation \eqref{eq:correlation_mat}, we can access the time-varying cross-covariances. These cross-covariances play a critical role in our channel estimation procedure.
\begin{remark}(Special cases: uniform angles and stationary environments)
 Define $\tau\triangleq t_2-t_1$ and $\mu \triangleq k-l$. If the tagged user and environment are stationary, i.e., the parameters $\nu(t)=\nu, \gamma(t)=\gamma, \theta_{\rm AoA}(t)=\theta_{\rm AoA}, \theta_{\rm AoD}(t)=\theta_{\rm AoD}$ are fixed and also we have uniform distribution for AoAs and AoD (isotropic scatterers), then, we have a simple closed-form formula for the $(k,l)$ correlation element as follows:
 
\begin{align}
   \rho_{k,l}(\tau,\mu)=J_0\left(-\tfrac{2\pi}{\lambda}\tau\nu\right)J_0\left(-\tfrac{2\pi}{\lambda}d \mu\right),
\end{align}
 
 which aligns with the well-known Jakes model in \cite[Eq. 4]{baddour2004accurate}. 
\end{remark}

\subsection{Uplink Pilot Signal Model}
\begin{figure}
    \centering
\includegraphics[scale=.2]{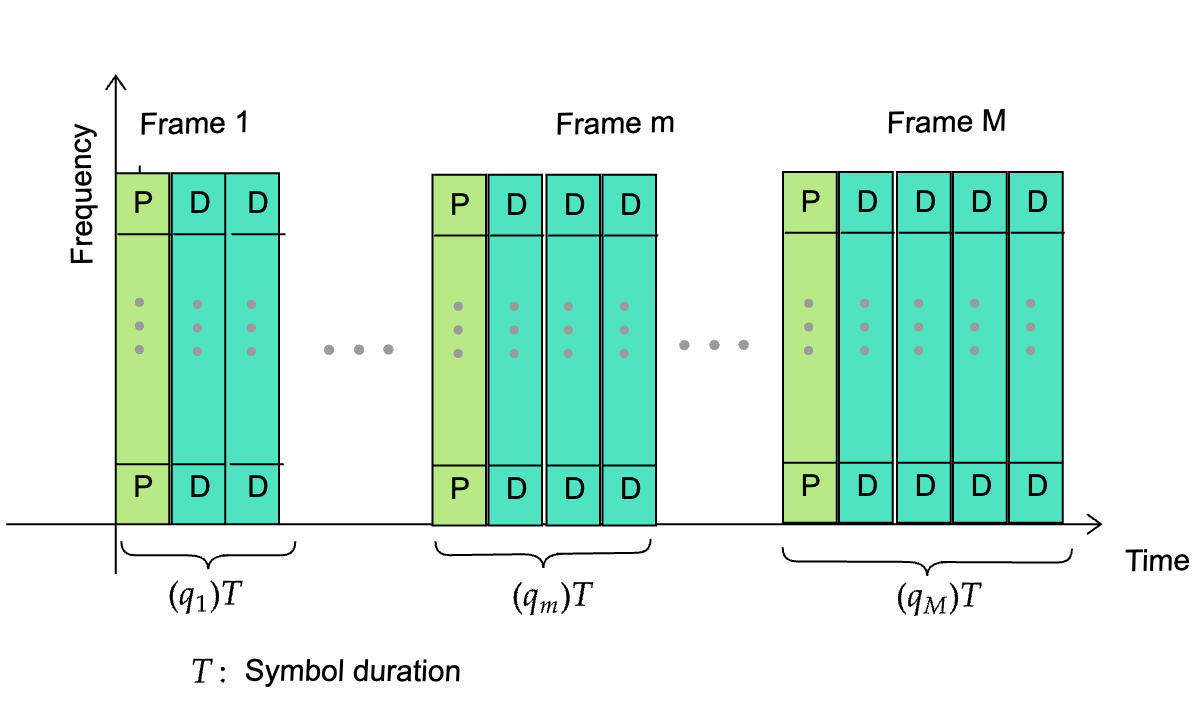}
    \caption{A schematic figure for multi-frame data transmission. $q_1,..., q_M$ specifies the length of frames $1,..., M$. Each time slot contains $F$ symbols in frequency domain. The first time slot of each frame is considered for pilot transmission (shown by P) and the rest is for data transmission (shown by D). 
}
    \label{fig:framework}
\end{figure}
We consider $M$ frames $m=1,..., M$ in which frame $m$ consists of $q_m$ time slots. The first time slot of each frame is devoted to pilot transmission and the rest is employed to transmit data as shown in Figure \ref{fig:framework}.   
Within frame $m$ out of the total of $M$ frames, each user sends $\tau_p$ pilot symbols in the first time slot,
followed by $q_m-1$ data time slots, each of which contains $F$ data symbols according to Figure \ref{fig:framework}.
Each symbol is transmitted within a duration of $T$ which is assumed to be one without loss of generality. By defining $\delta_m\triangleq\sum_{l=1}^m q_l +1$, the total duration is $\delta_M-1=\sum_{l=1}^M q_l$.
The times slots of frame $m$ are from $\delta_{m-1}$ to $\delta_{m}-1$. We consider a scenario involving $K$ single-antenna \acp{UE} and a \ac{BS} equipped with an \ac{ULA} consisting of $N_r$ antennas.
User-$k$ transmits each of the $F$ pilot symbols with transmit power $P_{{\rm p}, k}$, and each data
symbol in slot-$i$ with transmit power $P_{{\rm d},k},~k=1 \ldots K$.
Assuming that the coherence bandwidth accommodates at least $\tau_p$ pilot symbols,
this system allows one to create $\tau_p$ orthogonal pilot sequences collected in:

\begin{align}
\mathbf{s}&\triangleq \left[s(1),...,s({\tau_{\rm p}})\right]^{\rm T } \in \mathbb{C}^{{\tau_{\rm p} \times 1}},
\end{align}
whose elements satisfy 
$|s(j)|^2 = 1, s(j)s(k)=0, \forall j,k=1,..., \tau_p$.
Focusing on the received pilot signal from the tagged user at the \ac{BS},
the received pilot signal takes the following form:
\begin{align}\label{eqn:received_training_seq}
\mx{Y}_p(i)
&=
\alpha\sqrt{P_{\rm p}} \mx{h}(i) \mx{s}^{\rm T } +\mathbf{N}_{\rm p}(i) ~~ \in \mathbb{C}^{N_r \times \tau_{\rm p}},
\end{align}
where 
$\mx{h}(i) \in \mathbb{C}^{N_r \times 1 }$ is the channel vector at time slot $i$ with mean $\overline{\mx{h}}(i)$ and covariance matrix $\mx{C}_{\mx{h}}(i) \in \mathbb{C}^{N_r \times N_r}$. 
Furthermore, $\alpha\triangleq\alpha_1$ denotes
large scale fading, $P_{\rm p}$ denotes the pilot power of the tagged user at each pilot time slot,
and $\mathbf{N}_{\rm p}(i)\in \mathbb{C}^{N_r \times \tau_{\rm p}}$
is the 
\ac{AWGN} with element-wise variance $\sigma_{\rm p}^2$. It is beneficial to write equation \eqref{eqn:received_training_seq} in a matrix-vector form as follows:
\begin{align}
    \mx{y}_p(i)\triangleq\textbf{vec}\left(\mathbf{Y}_p(i)\right)=\alpha\sqrt{P_{\rm p}} \overline{\mx{S}}\mathbf{h}(i) +\mx{n}_{\rm p}(i) ~\in \mathbb{C}^{\tau_{\rm p} N_r \times 1},
\end{align}
where
$\mathbf{ y}_p(i)$, $\mx{n}_{\rm p}(i)\triangleq{\rm vec}(\mathbf{ N}_p(i)) \in \mathbb{C}^{\tau_{\rm p} N_r \times 1}$
and
$\overline{\mx{S}} \triangleq \mathbf{I}_{N_r} \otimes \mx{s}\in \mathbb{C}^{\tau_{\rm p} N_r \times N_r}$ is such that $\overline{\mx{S}}^{\rm H}\overline{\mx{S}}=\tau_{\rm p}\mx{I}_{N_r}$. 
We also assume that the pilot power corresponding to each pilot time slot is obtained as $P_{\rm p}=\tfrac{{P_{\rm p}}_{\max}}{M}$, where ${P_{\rm p}}_{\max}$ denotes the maximum (total) pilot power of the tagged user.

\subsection{Data Signal Model}
At data time slot $i$ of frame $m$, the received signal at the \ac{BS} can be stated as follows:

\begin{align}
&\mathbf{y}(i)=   \underbrace{{\alpha} \mx{h}(i) \sqrt{P_{\rm d}} {x}(i)}_{\text{tagged user}}\nonumber\\
&+ \underbrace{\sum_{k=2}^K \alpha_{k} \mx{h}_k(i) \sqrt{P_{{\rm d},k}} {x}_{k}(i)}_{\text{co-scheduled \ac{MU-MIMO} users}}
+\mathbf{n}_{d}(i)\in\mathbb{C}^{N_r \times 1},
\label{eq:data_measurements}
\end{align}
\noindent where $\mathbf{y}(i)\in \mathbb{C}^{N_r \times 1}$;
and ${x}_{k}(i)\in\mathbb{C}^{1\times 1}$ denotes the transmitted data symbol of user $k$
at time slot $i$ of $m$-th frame with transmit data power $P_{{\rm d},k}$. Here, the data power of each data time slot is obtained as $P_{{\rm d},k}=\tfrac{{P_{{\rm d},k}}_{\max}}{\delta_M -1-M}$, where ${P_{{\rm d}, k}}_{\max}$ denotes the maximum data power of user $k$.
Furthermore, $\mathbf{n}_{d}(i)~\sim \mathcal{CN}\left(\mx{0},\sigma_{\rm d}^2 \mathbf{I}_{N_r}\right)$
is the \ac{AWGN} at the receiver. The sum of the maximum pilot and data powers should not exceed the total power budget, which is denoted by $P_{\rm tot}$.

\section{Proposed Scheme}
\label{Sec:Proposed_scheme}
In this section, based on the temporal correlation information obtained in Proposition \ref{prop.rician}, we first characterize the \ac{LMMSE} estimate for the time-varying Rician aging channel as a function of frame size, number of frames and pilot and data powers in Section \ref{sec:channel_estimnate}. Next, we calculate the covariance matrix of the channel estimate as a function of the aforementioned parameters, which provides the required part for the subsequent sections. In Section \ref{sec:data_estimate}, based on the channel estimates in \ref{sec:channel_estimnate}, we provide an optimal \ac{MMSE} receiver combiner in order to estimate the date message of the tagged user exploiting the \ac{AR} channel information in previous times. Next, in Section \ref{Sec:SINR}, based on channel estimate and data estimate, we calculate the instantaneous slot-by-slot \ac{SINR} of the tagged user over fast fading Rician non-stationary channels in Theorem \ref{thm.Instantanous_SINR}. The instantaneous \ac{SINR} depends on the random channel estimates obtained in Section \ref{sec:channel_estimnate}, and is a random variable. Next, we state Theorem \ref{thm.stiel} in this section that provides a deterministic expression for the instantaneous \ac{SINR}, named deterministic equivalent \ac{SINR}. This deterministic equivalent \ac{SINR} relies solely on temporal correlation details of the Rician channel, independent of channel estimates or measurements. It forms the basis for our analysis in Section \ref{Sec:proposed_strategy}.

\subsection{Linear \ac{MMSE} Channel Estimation}\label{sec:channel_estimnate}
We assume that in each data slot $i$ of frame $m$, the \ac{BS} utilizes the pilot signals of the current, previous and next frames to estimate the channel in data time slots of the current frame. It's important to highlight that, in the initial frame, we solely take into account one pilot before and two pilots after. Conversely, in the last frame, our consideration includes two pilots before. According to \eqref{eqn:received_training_seq}, the observed measurements at the pilot time slots corresponding to frames $m-1$, $m$ and $m+1$ are given by:
\begin{align}\label{eqn:measurements_pilot}
&\mx{Y}\left(\delta_{m-2}\right)=
\alpha_1 \sqrt{P_{p}}\mx{h}\left(\delta_{m-2}\right) \mathbf{s}^{\rm T } +\mathbf{N}\left(\delta_{m-2}\right),\nonumber\\
&\mx{Y}\left(\delta_{m-1}\right)=
\alpha_1 \sqrt{P_{p}}\mx{h}\left(\delta_{m-1}\right) \mathbf{s}^{\rm T } +\mathbf{N}\left(\delta_{m-1}\right),\nonumber\\
&\mx{Y}\left(\delta_{m}\right)=
\alpha_1 \sqrt{P_{p}}\mx{h}\left(\delta_{m}\right) \mathbf{s}^{\rm T } +\mathbf{N}\left(\delta_{m}\right).
\end{align}
    

By stacking the above relations in matrix-vector form, we have that
\begin{align}\label{eqn:measurements_pilot_vectorized}
   & {\mx{y}}_{\rm p}\triangleq \begin{bmatrix}
       \textbf{vec}(\mx{Y}(\delta_{m-2}))\\
       \textbf{vec}(\mx{Y}(\delta_{m-1}))\\
       \textbf{vec}(\mx{Y}(\delta_{m}))
   \end{bmatrix}=
   &\widetilde{\mx{S}}{\mx{h}_p}+{\bs{\epsilon}_{\rm p}}\in\mathbb{C}^{3N_r\tau_{\rm p} \times 1},
\end{align}
where ${\mx{h}_p}=\begin{bmatrix}
       \mx{h}(\delta_{m-2})\\
       \mx{h}(\delta_{m-1})\\
       \mx{h}(\delta_{m})
   \end{bmatrix}\in\mathbb{C}^{3N_r\times 1}$, $\widetilde{\mx{S}}=\mx{I}_{3N_r}\otimes \mx{s}$ and ${\bs{\epsilon}_{\rm p}}\sim \mathcal{CN}(\mx{0},\sigma^2_{\rm p} \mx{I}_{3N_r\tau_{\rm p}})$. Note that as a convention, we assumed that $q_l=0~\forall l\le 0$.  

Having the measurements in pilot time slots at frames $m-1$, $m$, $m+1$, we can estimate the channel at data time slot $i$ in frame $m$ as is stated in the following lemma.  

\begin{lem}
\label{lem:mmsechannel}	
The \ac{LMMSE} channel estimate at data time slot $i$ of the $m$-th frame, which approximates the \ac{AR} fast fading channel in time slot $i$,
based on the received measurements at pilot time slots of the previous and next frames is given by:
\begin{align}
\label{eq:hmmse}
&\widehat{\mx{h}}(\mx{q},i)=\nonumber\\
&\tfrac{1}{\alpha_1\sqrt{P_{\rm p}}\tau_{\rm p}} \mx{E}(\mx{q},i) \left(\mx{M}_m(\mx{q})+\tfrac{\sigma^2_{\rm p}}{\alpha_1^2 P_{\rm p} \tau_{\rm p}}\mx{I}_{3N_r}\right)^{-1}\widetilde{\mx{S}}^{\rm H}\widetilde{\mx{y}}_{\rm p}+\overline{\mx{h}}(i),
\end{align}
where, for $\forall i=1,...,\delta_M-1, \forall m=1,.., M$, 
$\mx{E}_m(\mx{q},i)$ and $ \mx{M}_m(\mx{q})$ are 
\begin{align}
&    \mx{E}_m(\mx{q},i)\triangleq
{\begin{bmatrix}
    \mx{C}_{\mx{h}}(\delta_{m-2},i)&\mx{C}_{\mx{h}}(\delta_{m-1},i)&\mx{C}_{\mx{h}}(\delta_{m}, i) 
\end{bmatrix}};\label{eqn:E_m}\\
& \mx{M}_m(\mx{q})\triangleq  \label{eqn:M_m}
\\&{\begin{bmatrix}
    \mx{C}_{\mx{h}}(\delta_{m-2})&\mx{C}_{\mx{h}}(\delta_{m-2},\delta_{m-1})&\mx{C}_{\mx{h}}(\delta_{m-2},\delta_{m})\\
    \mx{C}_{\mx{h}}(\delta_{m-1},\delta_{m-2})&\mx{C}_{\mx{h}}(\delta_{m-1})&\mx{C}_{\mx{h}}(\delta_{m-1},\delta_{m})\\
    \mx{C}_{\mx{h}}(\delta_{m}, \delta_{m-2})&\mx{C}_{\mx{h}}(\delta_{m},\delta_{m-1})&\mx{C}_{\mx{h}}(\delta_{m})
    \end{bmatrix}},\nonumber
\end{align} 
and $\widetilde{\mx{y}}_{\rm p}\triangleq \mx{y}_{\rm p}-\alpha_1\sqrt{P}_p \widetilde{\mx{S}}(\overline{\mx{h}}(i)\otimes \mx{1}_{3})$.
\end{lem}
Proof. See Appendix \ref{proof.channel estimate}.
\begin{remark}(Special cases)
When all frames have the same size, while lacking a \ac{LoS} component in the channels from users to the \ac{BS}, the estimated channel in \eqref{eq:hmmse} aligns with the findings obtained in \cite[Eq. 10]{Fodor:23}.
\end{remark}
\begin{remark}(Optimality)
   If the channel follows Gaussian distribution, the \ac{LMMSE} estimate \eqref{eq:hmmse} coincides with the \ac{MMSE} estimate and is optimal in terms of minimizing \ac{MSE}. Alternatively, when the channel exhibits sparse representations in the angular domain, compressed sensing methods can be employed. These methods exploit the temporal prior information to achieve optimal channel estimation (see e.g., \cite{daei2019exploiting,daei2019living,daei2019error} for more details).
\end{remark}

%
%
According to Lemma \ref{lem:mmsechannel}, the \ac{LMMSE} estimate of the channel can also be stated as:
\begin{align}
\nonumber
&\widehat{\mx{h}}(\mx{q},i)=\mx{E}_m(\mx{q},i)\nonumber\\
&  \left(\mx{M}_m(\mx{q})+\tfrac{\sigma^2_{\rm p}}{\alpha^2 P_{\rm p} \tau_{\rm p}}\mx{I}_{3N_r}\right)^{-1}\left(\mx{h}_p-\overline{\mx{h}}(i)\otimes \mx{1}_{3}+{\bs{\epsilon}}\right)+\overline{\mx{h}}(i),
\label{eq:MMSEt}
\end{align}
where ${\bs{\epsilon}}\sim \mathcal{CN}(\mx{0},\tfrac{\sigma^2_{\rm p}}{\alpha^2 \tau_{\rm p} P_{\rm p}}\mx{I}_{3N_r})$
\begin{cor}
\label{cor:rmmse}	

The covariance matrix of the estimated channel $\widehat{\mx{h}}(\mx{q}, i)$ in Lemma \ref{lem:mmsechannel} is given by:
\begin{align}
\label{eq:rmmse}
\mx{C}_{\widehat{\mx{h}}}(i)
=& \mx{E}_m(\mx{q}, i) \left(\mx{M}_m(\mx{q})+\tfrac{\sigma^2_{\rm p}}{\tau_{\rm p} \alpha^2 P_{\rm p}}\mx{I}_{3N_r}\right)^{-1} \mx{E}_m^{\rm H}(\mx{q}, i).
\end{align}

\end{cor}

In what follows, we use the latter channel covariance matrix provided in \eqref{eq:rmmse} to calculate the optimum \ac{MMSE} receiver at the BS.
\subsection{Optimal \ac{MMSE} Receiver for {AR} Aging Channels}\label{sec:data_estimate}
In this section, we aim to calculate the optimum receiver $\mx{g}\in\mathbb{C}^{1\times N_r}$ that estimates the transmitted data symbol of the tagged user, which is denoted by ${x}$. Without loss of generality, we assume that the transmitted symbol of the tagged user has zero mean with unit variance.
Specifically, the \ac{BS} estimates the transmitted symbol of the tagged user in slot $i$ based on two prior information with different accuracy levels provided at $i$ and $i_{\rm p}$. Note that $i_{\rm p}< i$ can be any arbitrary time slot. Both of these time slots can provide beneficial and different information about the real channel at time $i$. In our \ac{MMSE} receiver, we will take into account both of them.
The following vector collects the prior information at time instances $i$ and $i_{\rm p}$:
\begin{align}
   \bs{\zeta}(i)\triangleq \begin{bmatrix}
       \widehat{\mx{h}}(\mx{q},i)\\
       \widehat{\mx{h}}(\mx{q},i_{\rm p})
   \end{bmatrix}\in\mathbb{C}^{2N_r\times 1}. 
\end{align}
Then, the optimum receiver is expressed as:
\begin{align}
   \mx{g}^{\star}(\mx{q},i)\triangleq \mathop{\arg\min}_{\mx{g}\in\mathbb{C}^{1\times N_r}} \mathds{E}_{{x}|\bs{\zeta}(i)} [|\mx{g}\mx{y}-{x}| ].
\end{align}
The solution of the latter optimization problem is given by:

\begin{align}\label{eq:G^star}
     \mx{g}^{\star}(\mx{q},i)=\alpha_1\sqrt{P_{\rm d}}\mx{z}_1^{\rm H} \mx{F}(i)^{-1},
\end{align}
where 
\begin{align}
    &\mx{F}(i)\triangleq\sum_{k=1}^K \alpha_k^2P_{{\rm d}, k}\mx{D}_k+\sigma_{\rm d}^2\mx{I}_{N_r},\mx{D}_k\triangleq\mx{Q}_k+ \mx{z}_k\mx{z}_k^{\rm H},\\
    &\mx{z}_k\triangleq \overline{\mx{h}}_k(i)+\bs{\Psi}_k \widetilde{\bs{\zeta}}_k(i),\bs{\zeta}_k(i)\triangleq\begin{bmatrix}
        &\widehat{\mx{h}}_k(i)\\
        &\widehat{\mx{h}}_k(i_{\rm p})
    \end{bmatrix}\in\mathbb{C}^{2 N_r \times 1},\\
    &\widetilde{\bs{\zeta}}_k(i)\triangleq\bs{\zeta}_k(i)-\overline{\mx{h}}(i)\otimes \mx{1}_{2},\\
    &\bs{\Psi}_k\triangleq\begin{bmatrix}
        \mx{C}_{\widehat{\mx{h}}_k(\mx{q},i)} & \mx{A}(i_{\rm p})\mx{C}_{\widehat{\mx{h}}_k(\mx{q},i_{\rm p})}\\
    \end{bmatrix}\nonumber\\
    &\begin{bmatrix}
         \mx{C}_{\widehat{\mx{h}}_k(\mx{q},i)}&\mx{A}(i_{\rm p})\mx{C}_{\widehat{\mx{h}}_k(\mx{q},i_{\rm p})}\\
         \mx{C}_{\widehat{\mx{h}}_k(\mx{q},i_{\rm p})}\mx{A}(i_{\rm p})^{H}&\mx{C}_{\widehat{\mx{h}}_k(\mx{q},i_{\rm p})}
    \end{bmatrix},\\
    &\mx{Q}_k\triangleq\mx{C}_{\mx{h}_k(i)}-\bs{\Psi}_k  \begin{bmatrix}
         \mx{C}_{\widehat{\mx{h}}_k(\mx{q},i)} \\ \mx{C}_{\widehat{\mx{h}}_k(\mx{q},i_{\rm p})} \mx{A}(i_{\rm p})^{\rm H}
     \end{bmatrix}.
\end{align}
 
 This finding extends the result presented in \cite[Eq. 24]{Fodor:2021} to the Rician channel, incorporating a significant \ac{LoS} component alongside the \ac{NLoS} components. The proof follows a similar approach.

\subsection{{SINR} and {SE} Calculations}
\label{Sec:SINR}
In this section, we first calculate the instantaneous \ac{SINR} based on channel and data estimates provided in Sections \ref{sec:channel_estimnate}
and \ref{sec:data_estimate}. Then, based on the instantaneous \ac{SINR}, we obtain the random \ac{SE}, where its randomness comes from the stochastic nature of channel estimates. The following theorem provides a closed-form expression for the instantaneous \ac{SINR} that is calculated at the \ac{BS} side.
\begin{thm}\label{thm.Instantanous_SINR}
Assume that there is some prior information about the channel at time $i$ collected in $\bs{\zeta}(i)$. Also, the receiver employs the \ac{MMSE} combiner to estimate the data symbols of the tagged user at time slot $i$. Then, the instantaneous \ac{SINR} of the data symbol of the tagged user at time slot $i$ is obtained as:
\begin{align}\label{eq:Instantanous_SINR_simplified}
     &\gamma(\mx{q},i,\zeta(i))= \alpha^2 P_{\rm d} \mx{z}_1^{\rm H}\left(\mx{F}(i)- \alpha_1^2 P_{\rm d} \mx{z}_1 \mx{z}_1^{\rm H}\right)^{-1} \mx{z}_1\nonumber\\
     &~~~~~~~~~~~~~~~~~~\alpha_1^2 P_{\rm d}  \mx{z}_1^{\rm H} \left( \mx{F}_1^{-1}(i)\right) \mx{z}_1,
     \end{align}
     where 
\begin{align}\label{eq:F1}
&\mx{F}_1\triangleq \mx{F}_1(i)\triangleq \mx{F}(i)-\alpha_1^2 {P_{\rm d}} \mx{z}_1\mx{z}_1^{\rm H}.
\end{align}

\end{thm}
%
Proof. See Appendix \ref{proof.thm.SINR}.

The latter result immediately leads to finding the random \ac{SE} as follows:
\begin{align}\label{eq:random_SE}
\textup{SE}\Big(\mx{q},i,
\bs{\zeta}(i)\Big)\triangleq\log\Big(1+\gamma(\mx{q},i,\bs{\zeta}(i))\Big),
\end{align}
which is a random variable. In the next theorem, leveraging concentration inequality results from random matrix theory tools as detailed in \cite{bai2010spectral}, we present a deterministic equivalent expression for the \ac{SE}. This expression serves as a reliable approximation for the average \ac{SE}, when the number of \ac{BS} antennas becomes large.

\begin{thm}\label{thm.stiel}
 Define $\mx{S}\triangleq \sum_{k=1}^K \alpha_k^2 P_{{\rm d}, k}\mx{Q}_k$ and $\rho_{\rm d}\triangleq \sigma_{\rm d}^2$. When $N_r$ is sufficiently large ($N_r\rightarrow \infty$), the instantaneous \ac{SE} provided in \eqref{eq:random_SE} is concentrated around a deterministic expression given by:
\begin{align}\label{eq:DSE}
  &{\textup{SE}}^{\circ}(\mx{q}, i)\triangleq\nonumber\\
  &\log\left(1+\left\langle \mx{R}_{\mx{z}_1}, \left(\sum_{k=2}^K \tfrac{\alpha_k^2 P_{{\rm d}, k}\mx{R}_{\mx{z}_k}}{1+m_{\mx{B}_{k}}(\rho_{\rm d})}+\mx{S}+\rho_{\rm d} \mx{I}_{N_r}\right)^{-1} \right\rangle \right),
\end{align}
where $m_{\mx{B}_{(2)}}(\rho_{\rm d}),\ldots,m_{\mx{B}_{k}}(\rho_{\rm d})$ are the solution of the following system of equations
for $k=2,\ldots,K$:
\begin{align}
   &m_{\mx{B}_{k}}(\rho_{\rm d}) =\left\langle \mx{R}_{\mx{z}_k},\left(\sum_{l=2}^K\tfrac{\mx{R}_{\mx{z}_l}}{1+m_{\mx{B}_{l}}(\rho_{\rm d})}+\mx{S}+\rho_{\rm d} \mx{I}_{N_r}\right)^{-1}\right\rangle\nonumber
\end{align}
\end{thm}
Proof. See Appendix \ref{proof.thm.stiel}.

We call the deterministic expression \eqref{eq:DSE} the deterministic equivalent \ac{SE}. In the next section, we exploit this useful expression in order to propose optimal frame design and power control.

\section{Proposed Strategy for Frame and Power Design}
\label{Sec:proposed_strategy}
In this section, we formulate an optimization problem to design the optimal values for the number of frames (denoted by $M^{\star}$), the frame size (denoted by $\mx{q}_m^{\star}$), maximum pilot power (${P_{\rm p}}_{\max}^{\star}$) and maximum data power ${P_{\rm d}}_{\max}^{\star}$ under the condition that we have a limited total power budget (denoted by $P_{\rm tot}$).

The objective function that we intend to maximize is \ac{ASE}, which is obtained by taking the average of $\text{SE}^{\circ}(\mx{q}, i)$ over all data time slots of all frames defined as
\begin{align}\label{eq:deterministic ASE function}
\text{ASE}\triangleq\tfrac{\sum_{l=1}^{\delta_M -1}\text{SE}^{\circ}(\mx{q}, l)}{\delta_M -1}, ~~\delta_M>1.
\end{align}

The proposed optimization problem that finds the optimal values of $M$, $q_m$, ${P_{\rm p}}_{\max}$ and ${P_{\rm d}}_{\max}$ is provided below:
\begin{align}\label{eq:opt_problem}
\{\mx{q}_{\max}^\star,M_{\max}^{\star},& {P_{\rm p}}_{\max}^\star, {P_{\rm d}}_{\max}^\star\}={\arg\max}_{\mx{q},M, {P_{\rm p}}_{\max}, {P_{\rm d}}_{\max}} \text{ASE},\\ \nonumber
~~&{\rm s.t.}~~{P_{\rm p}}_{\max}+{P_{\rm d}}_{\max}\le P_{\rm tot},
\end{align}
where ${P_{\rm d}}_{\max}={P_{{\rm d},1}}_{\max}$ and ${P_{\rm p}}_{\max}={P_{{\rm p},1}}_{\max}$ denote the maximum data 
 and pilot power of the tagged user, respectively.
\begin{remark}(Critical factors in the proposed optimization problem \eqref{eq:opt_problem})
   In our numerical experiments, it has been observed that while the interference components (such as interference pilot and data power of other users and interference channel including Doppler frequency and path loss of other users) affect the resulting \ac{ASE} of the tagged user, they do not have any role in determining the optimal frame design ($\mx{q}^{\star}$ and $M^{\star}$). This feature enables our method to be employed at the transmitter side (user side) rather than the receiver side (\ac{BS} side). 
\end{remark}
\subsection{A Heuristic Algorithm for Optimal Frame and Power Design}
\label{Sec:Alg}
The optimization problem \eqref{eq:opt_problem} is a mixed-integer nonlinear problem that is in general \ac{NP}-hard. However, in this section, we provide a heuristic algorithm named OptResource that calculates the optimal values of frame size, number of frames and pilot and data powers. The pseudo code of this algorithm is provided in Algorithm \ref{alg.optweights}. For each number of frames and for each frame size, a projected gradient ascent is performed to find the optimal values of pilot and data powers (Lines 10 to 16). The projected gradient ascent consists of two steps: update step and projection. In the update step in Line 14, a regular gradient ascent is performed. The resulting variables after this update step might not be inside the feasible region (the sum of pilot and power must be less that the total power budget). The updated variables of power is then projected in Line 16 to the closest point in the feasible region. The algorithm scans through all possible values of $M$ and $q_m$ that lead to the maximum spectral efficiency. The final outputs of OptResource is $\mx{q}^\star, M^\star, {P_{\rm p}}_{\max}^{\star}, {P_{\rm d}}_{\max}^{\star}$. It is worth mentioning that the computational complexity of OptResource is $\mathcal{O}(\mx{q}_{\max} K N_r^3)$.

\alglanguage{pseudocode}
{\centering
 \resizebox{.5\textwidth}{!}{
\begin{minipage}{.7\textwidth}
\begin{algorithm}[H]
	\caption{Proposed algorithm for optimal frame and power design}\label{alg.optweights}
	\begin{algorithmic}[1]
		\Procedure {OptResource}{$\mx{P}_{\mx{h}}(t_1,t_2)$, $\rm Tol$, $\mx{q}_{\max}$, $M_{\max}$, ${\rm maxiter}$, $P_{\rm tot}$}
  \State Define  $\mx{w}\triangleq[{P_{\rm p}}_{\max},{P_{\rm d}}_{\max}]^{\rm T },$
  \State Choose a small constant for the steepest ascent step size $\mu_{\mx{w}}$.
  \State Objective function : ${\rm SE}(q_1,...,q_M,M,\mx{w})$ in \eqref{eq:deterministic ASE function}
  \State  Pick an initial point $\mx{w}^0$ for maximum pilot and data power that satisfies $\mx{1}^{\rm T }\mx{w}^0 \le P_{\rm tot}$ 
  \State $M=1, {\rm SE}^\star=0$
  \While{$M\le M_{\max}$}
\For{$i_1=1~{\rm to} \lceil\tfrac{q_{\max}}{M}\rceil $}

\State \vdots
\For{$i_M=1~{\rm to} \lceil \tfrac{q_{\max}}{M}\rceil $}

\State Objective function for power
	\State	$f( \mx{w})={\rm SE}(i_1,...,i_M,M,\mx{w})$   \label{step.costfun}

  \State $k\leftarrow 1$
        \While{$\|\mx{w}^{k}-\mx{w}^{k-1}\|_2 >\rm Tol$ and $|f(\mx{w}^{k})-f(\mx{w}^{k-1})|>\rm Tol$ and $k<{\rm maxiter}$} 

Compute ascent direction by calculating $\tfrac{\partial {\rm SE}}{\partial \mx{w}}$ at the point $\mx{w}^k$	\\
 Update according to
         $\widetilde{\mx{w}}^{k+1}\leftarrow \mx{w}^{k}+\mu_{\mx{w}} \tfrac{\partial {\rm SE}}{\partial \mx{w}} $\\
 Project to the closest point inside feasible region
 \\
$\mx{w}^{k+1}=\mathop{\arg\min}_{\mx{x}}\|\mx{x}-\widetilde{\mx{w}}^{k+1}\|_2~~{\rm s.t.} ~\mx{1}^{\rm T } \mx{x}\le P_{\rm tot}$
  \State $k\leftarrow k+1$

	\EndWhile
\State $\mx{w}^{\star} \leftarrow \mx{w}^k$
\State ${\rm SE}={\rm SE}(q_1,...,q_M,M,\mx{w}^{\star})$

\If{${\rm SE}> {{\rm SE}}^\star$}

\State ${{\rm SE}}^{\star}\leftarrow SE$
\State $\mx{q}^\star\leftarrow[i_1,,..., i_M]^{\rm T }$
\State $M^{\star}\leftarrow M$
  \EndIf	
  \EndFor
  
\State   \vdots

  \EndFor
  \State $M\leftarrow M+1$
  \EndWhile
\Statex $[{P_{\rm p}}_{\max}^\star , {P_{\rm d}}_{\max}^\star]^{\rm T }\leftarrow\mx{w}^\star$
		\EndProcedure
	\end{algorithmic}
 Outputs: $\mx{q}^\star, M^\star,{P_{\rm p}}_{\max}^\star , {P_{\rm d}}_{\max}^\star$
\end{algorithm}
\end{minipage}}}
\section{Numerical Results}
\label{Sec:simulations}
In this section, we examine some numerical experiments to evaluate the performance of our proposed method. 
In the first experiment, we consider $q_{\max}=12$ and $M_{\max}=4$ and compare four cases in Table \ref{table:comparison}: 1. use one frame with one pilot time slot and $11$ data time slots. 2. use $M=2$ frames with two pilot time slots at the first of each frame. 3. use $M=3$ time slots with three pilots. 4. use $M=4$ frames with four pilot time slots. Each of the four rows in Table \ref{table:comparison} corresponds to one experiment. The first three experiments are related to the stationary settings while the rest examine the performance of our method in non-stationary scenarios. We considered $K=2$ users and the first user is assumed to be the tagged user. The Doppler frequencies, pilot power, data power, Rician K factor, path loss and channel variances are shown respectively by ${f_d}_i, {P_{{\rm p},i}}_{\max}, {{P_{{\rm d},i}}}_{\max}, {K_f}_i, {\rm PL}_i, \sigma_{\mx{h}_i}^2$ where $i=1$ corresponds to the tagged user and $i=2$ shows the interference component. The optimal values for the number $M$ of frames, frame size $\mx{q}$ and their corresponding deterministic ASEs are shown by bold numbers. 

The pilot and data power in this experiment are assumed to be fixed. Note that \ac{SNR} is defined as ${\rm SNR}\triangleq10\log(\tfrac{P_{\rm d}}{\sigma_{\rm d}^2})-{\rm PL}_1$ where ${\rm PL}_1\triangleq20\log(\alpha_1)$ is the path loss of the channel corresponding to the tagged user. Note that the values of both the SNR and the path loss are shown in dB units. As it turns out from Table \ref{table:comparison}, in the first case (first four rows), the Doppler frequency of the tagged user is small and our proposed method suggests to use the maximum data possible slots with one frame while in the third case where the tagged user is fast, it suggests to use $4$ frames with optimal frame sizes $\mx{q}^{\star}=[3,3,3,2]$. Also, by comparing the first and second cases, we observe that high Rician K factor leads to higher SEs. We also observe from the last case that $\mx{q}^{\star}=[5,2]$ is optimal and it suggests that using all time slots does not necessarily lead to higher \ac{SE} values. We also observe that the main key factors in determining optimal frame design is Doppler, K factor of the tagged user and other parameters do not affect the optimal number of frames and frame size.

In the second experiment, we examined the effects of Doppler and Rician K factor in the optimal frame design shown in Figures
\ref{fig:fig3,4} and \ref{fig:fig5,6,7}
in the settings that we have one frame with $q_{\max}=24$. In the zero K factor setting in the left image of Figure \ref{fig:fig3,4}, as you observe, the optimal frame size is shifting to the right as the tagged user's speed decreases. It suggests that in high speed scenarios, use all the power budge in initial time slots as the channel is highly dynamic. In the right image of Figure \ref{fig:fig3,4}, the proposed method suggests to use more time slots in good \ac{LoS} conditions. The more \ac{LoS} power the channel has, the more \ac{SE} you get. It is also interesting to observe in the upper-right image of Figure \ref{fig:fig5,6,7} that even though the Doppler is high but in cases that the tagged channel has strong LoS condition, the proposed method suggests to use more time slots in different \ac{SNR} conditions. However, it experiences less \ac{ASE} compared to the lower Doppler frequency case  shown in the upper-left image of Figure \ref{fig:fig5,6,7}. We also provide a 3D plot in the lower image of Figure \ref{fig:fig5,6,7} where \ac{ASE} is drawn versus both Doppler and frame size in different \ac{SNR} conditions. 

In the third experiment shown in Figure \ref{fig:fig7,8,9}, we have performed some experiments to investigate the elements that have impacts on the optimal frame design and power. The left, middle and right images of Figure \ref{fig:fig7,8,9} highlight the fact that interference components such as path loss, Doppler frequency and power of other users can not affect the optimal frame size and optimal number of frames in the tagged user. As it turns out in these figures, while different Doppler, path loss and powers of interference can change the resulting deterministic \ac{ASE}, optimal frame size is not sensitive to these interference components. Here, ${r_p}_i=\tfrac{{{P_{{\rm p}, i}}}_{\max}}{{{P_{{\rm d}, i}}}_{\max}}$
specifies the distribution of pilot to data power for the tagged user ($i=1$) and interference ($i=2$). In the left image of Figure \ref{fig:fig10,11}, we observe that the power distribution between pilot and data (captured by $r_p$) can affect the optimal frame size.

The fourth experiment shows the results of the joint frame size and power optimization, and illustrates the effect of the interference power distribution (captured by ${r_p}_2$) on the optimal power, see the right image of Figure \ref{fig:fig10,11}. 
We observe in the right image of Figure \ref{fig:fig10,11} that while ${r_p}_2$ affects ${{P_{\rm p}}^\star_{\max}}$ and ${{P_{\rm d}}^{\star}_{\max}}$, it does not play any role in determining $\mx{q}^\star$ and $M^\star$. Additionally, the maximum deterministic \ac{ASE} occurs in the case of  $M^\star=2$, and $\mx{q}^{\star}=[3,2]$ in the settings provided in the caption of the right image of Figure \ref{fig:fig10,11}.
{\centering
\begin{table*}[t]
\caption{Performance of our proposed method in diverse non-stationary and stationary scenarios}
\centering
\label{table:comparison}
\tiny
\begin{tabular}{|p{.8cm}|p{.03cm}|p{.7cm}|p{.25cm}|p{.5cm}|p{.5cm}|p{.65cm}|p{.65cm}|p{.1cm}|p{1cm}|p{.3cm}|p{1cm}|p{.4cm}|p{.21cm}|p{.22cm}|p{0.75cm}|p{0.2cm}|}
    \hline
    
    ${\mx{q}^{\star}}$ &$M$ &SE& $ {\rm SNR}$&${P_{\rm p}}_{\max}$&${P_{\rm d}}_{\max}$&${{P_{\rm p}}_2}_{\max}$&${{P_{\rm d}}_2}_{\max}$&$N_r$&${f_d}_1$&${f_d}_2$&  ${K_1}_{\rm f}$ & ${K_2}_{\rm f}$ &${P_L}_1$&${P_L}_2$& $\sigma^2_{\mx{h}_1}$&$\sigma^2_{\mx{h}_2}$ \\
    \hline
    \hline
     $\mx{12}$ &$\mx{1}$&$\mx{13.3067}$      &$30$&$1$&$1$&$1$&$1$&$20$&$0.1$&$100$&$1$&$0$&$1$&$90$&$1$&$1$\\ \hline
    $[6,6]$ &$2$&$12.9756$    &$30$&$1$&$1$&$1$&$1$&$20$&$0.1$&$100$&$1$&$0$&$1$&$90$&$1$&$1$\\ \hline
    $[4,4,4]$ &$3$&$11.9469$  &$30$&$1$&$1$&$1$&$1$&$20$&$0.1$&$100$&$1$&$0$&$1$&$90$&$1$&$1$\\ \hline
    $[3,3,3,3]$&$4$&$10.7511$ &$30$&$1$&$1$&$1$&$1$&$20$&$0.1$&$100$&$1$&$0$&$1$&$90$&$1$&$1$\\ \hline\hline
    $12$ &$1$&$3.8722$        &$0$&$.1$&$.1$&$.1$&$.1$&$20$&$1$&$10$&$1$&$0$&$1$&$90$&$1$&$1$\\ \hline
    $\mx{[6,6]}$ &$\mx{2}$&$\mx{4.3383}$    &$0$&$.1$&$.1$&$.1$&$.1$&$20$&$1$&$10$&$0.1$&$0$&$0$&$90$&$1$&$1$\\ \hline
    $[4,4,4]$ &$3$&$4.0842$  &$0$&$.1$&$.1$&$.1$&$.1$&$20$&$1$&$10$&$0.1$&$0$&$1$&$90$&$1$&$1$\\ \hline
    $[3,3,3,3]$&$4$&$3.7122$ &$0$&$.1$&$.1$&$.1$&$.1$&$20$&$1$&$10$&$0.1$&$0$&$1$&$90$&$1$&$1$\\ \hline\hline
    $2$ &$1$&$1.5806$        &$30$&$.1$&$.1$&$.1$&$.1$&$20$&$100$&$0.01$&$0 $&$1$&$1$&$90$&$1$&$1$\\ \hline
    $[3,2]$ &$2$&$2.0004$    &$30$&$.1$&$.1$&$.1$&$.1$&$20$&$100$&$0.01$&$0$&$1$&$1$&$90$&$1$&$1$\\ \hline
    $[3,3,2]$ &$3$&$2.1054$  &$30$&$.1$&$.1$&$.1$&$.1$&$20$&$100$&$0.01$&$0$&$1$&$1$&$90$&$1$&$1$\\ \hline
    $\mx{[3,3,3,2]}$&$\mx{4}$&$\mx{2.1531}$ &$30$&$.1$&$.1$&$.1$&$.1$&$20$&$100$&$0.01$&$0$&$1$&$1$&$90$&$1$&$1$\\ \hline\hline
    $\mx{6}$ &$\mx{1}$&$\mx{9.1511}$        &$50$&$1$&$1$&$1$&$1$&$20$&$0.1 t$&$0.1t$&$0.1 t$&$0.1 t$&$50$&$50$&$\tfrac{10}{t}$&$\tfrac{10}{t}$\\ \hline
    ${[6,4]}$ &${2}$&${8.773}$    &$50$&$1$&$1$&$1$&$1$&$20$&$0.1 t$&$0.1t$&$0.1 t$&$0.1 t$&$50$&$50$&$\tfrac{10}{t}$&$\tfrac{10}{t}$\\ \hline
    $[4,4,4]$ &$3$&$8.2585$  &$50$&$1$&$1$&$1$&$1$&$20$&$0.1 t$&$0.1t$&$0.1 t$&$0.1 t$&$50$&$50$&$\tfrac{10}{t}$&$\tfrac{10}{t}$\\ \hline
    $[3,3,3,3]$&$4$&$7.5917$ &$50$&$1$&$1$&$1$&$1$&$20$&$0.1 t$&$0.1t$&$0.1 t$&$0.1 t$&$50$&$50$&$\tfrac{10}{t}$&$\tfrac{10}{t}$\\ \hline\hline
    $\mx{10}$ &$\mx{1}$&$\mx{3.671}$        &$10$&$10$&$1$&$10$&$1$&$20$&$10(12-t)$&$10 t$&$10(12-t)$&$10 t$&$1$&$1$&$\tfrac{1}{10(12-t)}$&$\tfrac{1}{10t}$\\ \hline
    $[6,6]$ &$2$&$3.3593$    &$10$&$10$&$1$&$10$&$1$&$20$&$10(12-t)$&$10 t$&$10(12-t)$&$10 t$&$1$&$1$&$\tfrac{1}{10(12-t)}$&$\tfrac{1}{10t}$\\ \hline
    $[4,4,2]$ &$3$&$3.0668$  &$10$&$10$&$1$&$10$&$1$&$20$&$10(12-t)$&$10 t$&$10(12-t)$&$10 t$&$1$&$1$&$\tfrac{1}{10(12-t)}$&$\tfrac{1}{10t}$\\ \hline
    $[3,3,3,2]$&$4$&$2.7636$ &$10$&$10$&$1$&$10$&$1$&$20$&$10(12-t)$&$10 t$&$10(12-t)$&$10 t$&$1$&$1$&$\tfrac{1}{10(12-t)}$&$\tfrac{1}{10t}$\\ \hline\hline
    $6$ &$1$&$0.8886$        &$0$&$1$&$1$&$1$&$1$&$20$&$1(12-t)$&$10 t$&$0$&$0$&$50$&$100$&$0$&$0$\\ \hline
    $[6,6]$ &$2$&$1.7832$    &$0$&$1$&$1$&$1$&$1$&$20$&$1(12-t)$&$10 t$&$0$&$0$&$50$&$100$&$0$&$0$\\ \hline
    $\mx{[4,4,4]}$ &$\mx{3}$&$\mx{1.9684}$  &$0$&$1$&$1$&$1$&$1$&$20$&$1(12-t)$&$10 t$&$0$&$0$&$50$&$100$&$0$&$0$\\ \hline
    $[3,3,3,3]$&$4$&$1.9231$ &$0$&$1$&$1$&$1$&$1$&$20$&$1(12-t)$&$10 t$&$0$&$0$&$50$&$100$&$0$&$0$\\ \hline\hline
    $5$ &$1$&$2.9595$        &$0$&$0.1$&$0.1$&$0.1$&$0.1$&$20$&$5 t$&$10 $&$\tfrac{t}{50}$&$\tfrac{12-t}{12}$&$0$&$90$&$1$&$1$\\ \hline
    $\mx{[5,2]}$ &$\mx{2}$&$\mx{3.0616}$    &$0$&$0.1$&$0.1$&$0.1$&$0.1$&$20$&$5 t$&$10$&$\tfrac{t}{50}$&$\tfrac{12-t}{12}$&$0$&$90$&$1$&$1$\\ \hline
    $[4,4,2]$ &$3$&$2.9445$  &$0$&$0.1$&$0.1$&$0.1$&$0.1$&$20$&$5 t$&$10$&$\tfrac{t}{50}$&$\tfrac{12-t}{12}$&$0$&$90$&$1$&$1$\\ \hline
    $[3,3,3,2]$&$4$&$2.7627$ &$0$&$0.1$&$0.1$&$0.1$&$0.1$&$20$&$5 t$&$10$&$\tfrac{t}{50}$&$\tfrac{12-t}{12}$&$0$&$90$&$1$&$1$\\ \hline
\end{tabular}
\end{table*}
}
\begin{figure}
    \centering
    \includegraphics[scale=.7]{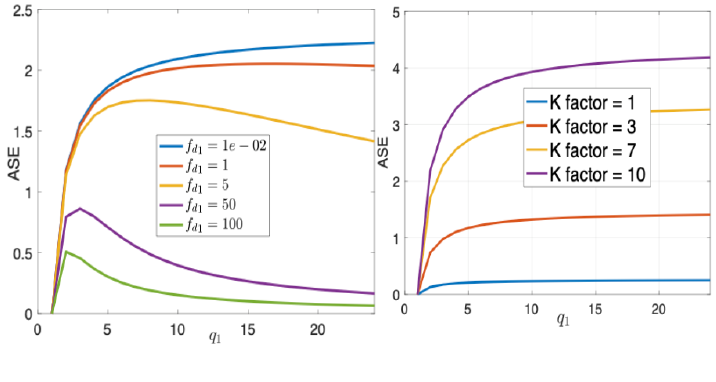}
    \caption{Left image: \ac{ASE} versus Doppler frequency  Right image:\ac{ASE} versus Rician K factor.  The parameters related to these two experiments are set as follows: $N_r=20, \mx{C}_{\mx{h}_1}=\mx{I}_{N_r}, {P_{\rm p}}_{\max}={P_{\rm d}}_{\max}=1, {\rm SNR}=0~dB, f_c=1000,{P_{\rm p}}_{\max}={P_{\rm d}}_{\max}=1, \tau_p=1,  {\rm PL}_1={\rm PL}_2=0, q_{\max}=24, M=1, {K_f}_2=0$. In the left image, ${K_f}_1=0$ and in the right image, $f_d=100$.}
    \label{fig:fig3,4}
\end{figure}

\begin{figure}
    \centering
    \includegraphics[scale=.6]{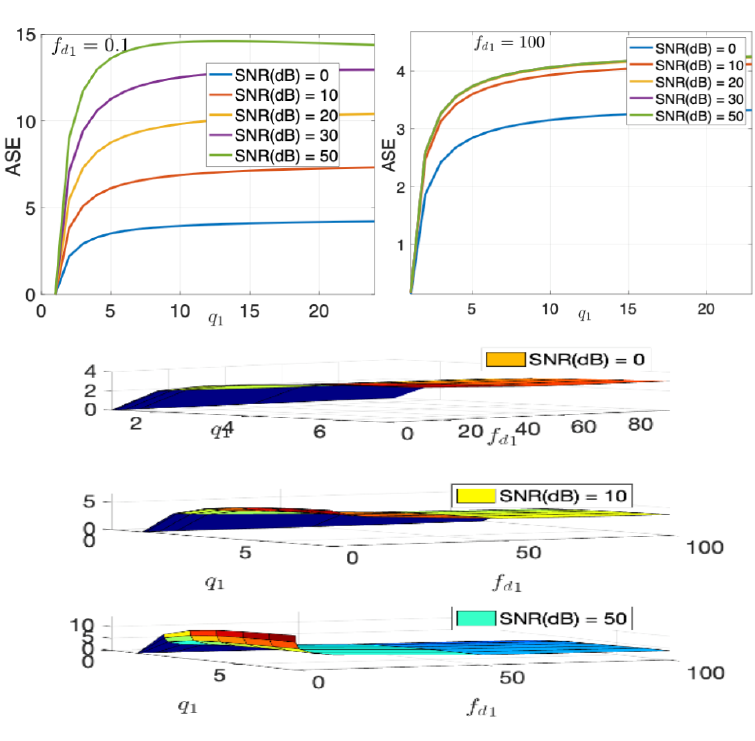}
    \caption{The upper left and right images show \ac{ASE} versus SNR in the case of ${f_d}_1=0.1$ and ${f_d}_1=100$, respectively. The bottom image
    shows \ac{ASE} versus both Doppler and \ac{SNR} in different time slots. The parameters related to these experiments are set as follows:
   ${f_d}_2=100, f_c=1000, {\rm PL}_1=0, {\rm PL}_2=90, q_{\max}=24, M=1, {K_f}_1=1, {K_f}_2=0, {P_{\rm p}}_{\max}={P_{\rm d}}_{\max}=1, N_r=20, \tau_p=1$.}
    \label{fig:fig5,6,7}
\end{figure}
\begin{figure}
    \centering
    \includegraphics[scale=.65]{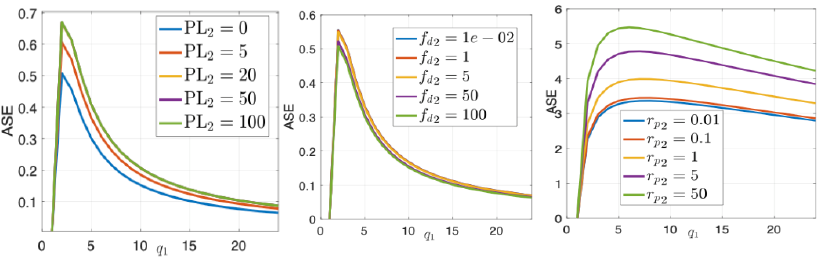}
    \caption{The three images, from left to right, depict the \ac{ASE} versus interference path loss, interference Doppler and interference power ratio under the following configurations: Left image: ${f_d}_1={f_d}_2=100, {\rm PL}_1=0, {P_{\rm p}}_{\max}={P_{\rm d}}_{\max}=1$, Middle image: ${f_d}_1=100, {\rm PL}_1={\rm PL}_2=0, {P_{\rm p}}_{\max}={P_{\rm d}}_{\max}=1$, Right image: ${f_d}_1=5, {f_d}_2=100,{\rm PL}_1={\rm PL}_2=0$. The remaining parameters for these experiments are consistent and set as follows: $q_{\max}=24, M=1, {K_f}_1=0, {K_f}_2=0, N_r=20, \tau_p=1, {\sigma_{\rm p}^2}_1={\sigma_{\rm p}^2}_2={\sigma_{\rm d}^2}_1=10^{-4}$.}
    \label{fig:fig7,8,9}
\end{figure}
\begin{figure}
    \centering
    \includegraphics[scale=.65]{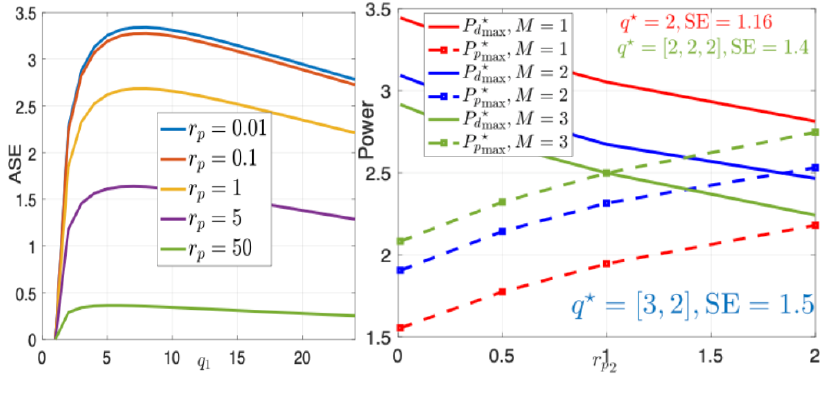}
    \caption{Left image: \ac{ASE} versus power ratio with parameters ${f_d}_1=5, {f_d}_2=100, {\rm PL}_1={\rm PL}_2=0, q_{\max}=24, M=1$. Right image:
    Pilot and data power distribution versus interference power with parameters ${f_d}_1=100, {f_d}_2=0.1, {\rm PL}_1=10, {\rm PL}_2=50, {K_f}_1=0, {K_f}_2=0, N_r=10, \tau_p=1, {\sigma_{\rm p}^2}_1=10^{-5}, {\sigma_{\rm p}^2}_2=10^{-6}, {\sigma_{\rm d}^2}=10^{-5}$. Also, the parameters that are similar in both experiments are as follows: $f_c=1000, {K_f}_1=0, {K_f}_2=0, N_r=20,\tau_p=1$.}
    \label{fig:fig10,11}
\end{figure}
\section{Conclusions}
\label{sec:conclusion}
This study investigated uplink communications systems operating over fast-fading Rician non-stationary channels that experience aging between subsequent pilot time slots. An analytical framework was proposed for determining the optimal parameters such as frame size, number of frames, and power distribution between data and pilot slots. We proposed an analytical framework to identify optimal parameters, including frame size, the number of frames, and power allocation between data and pilot slots. The proposed methodology revolves around optimizing a deterministic equivalent expression for spectral efficiency. Importantly, our optimization approach relied solely on the knowledge of the temporal dynamics of the channels, excluding the need for measurements or channel estimates. The core objective was to ascertain the optimal rate at which we should update our understanding of the channel's temporal behavior within dynamic environments. We also demonstrated that optimal frame design can be performed at the transmitter side, while optimal pilot and power configurations can be accomplished at the receiver side. Moreover, we proposed an efficient algorithm to find the optimal parameters.

Simulation results validated the efficacy of our methodology, highlighting its impact on optimizing pilot power, data power, frame duration, and the number of frames across diverse practical scenarios. This work contributes to the advancement of uplink communication systems by providing a novel approach that leverages predictive insights into channel behavior for improved performance in dynamic and aging channel environments.

\appendices
\section{Proof of Proposition \ref{prop.channel_model}}\label{proof.channel_model}
The channel at time $t+1$ is described based on mean and covariance. According to \eqref{eq:AR1}, the mean of the channel at time $t+1$ is $ \mathds{E}[\mx{h}(t+1)]= \overline{\mx{h}}(t+1)$,
which is a result of $\mathds{E}[\widetilde{\mx{h}}(t)]=\mathds{E}[\bs{\xi}(t+1)]=\mx{0}$.
Also, based on the \ac{AR} process in \eqref{eq:AR1}, the covariance matrix of the channel at time $t+1$ is obtained as follows:
\begin{align}\label{eq:relar1}
 & \mx{C}_{\mx{h}}(t+1)  =\mx{A}(t)\mx{C}_{\mx{h}}(t) \mx{A}^{\rm H}(t)+\bs{\Theta}(t+1) 
\end{align}
Furthermore, based on the definition of the cross-covariance matrix, we have:
\begin{align}
    \mathds{E}[\widetilde{\mx{h}}(t+1)\widetilde{\mx{h}}^{\rm H}(t)]=\mx{A}(t)\mx{C}_{\mx{h}}(t)=\mx{C}_{\mx{h}}(t+1,t)
\end{align}
which results in
\begin{align}\label{eq:relar2}
    \mx{A}(t)=\mx{C}_{\mx{h}}(t+1,t)\mx{C}^{-1}_{\mx{h}}(t)
\end{align}
By combining \eqref{eq:relar1} and \eqref{eq:relar2}, the covariance of \ac{AR} noise is characterized as below:
\begin{align}
    \bs{\Theta}(t+1)=\mx{C}_{\mx{h}}(t+1)-\mx{C}_{\mx{h}}(t+1,t)\mx{C}^{-1}_{\mx{h}}(t)\mx{C}_{\mx{h}}(t,t+1).
\end{align}

The last property that our model should satisfy is  what we call the correlation decaying property. In fact, the correlation 
between times $t+n$ and $t$ must be decaying as the integer distance $n$ increases. To prove this, we first write the \ac{AR} process based on our model for the normalized centered channel $\mx{h}^{\prime}(t)$ which is obtained as follows:
\begin{align}\label{eq:model_for_normalized_channel}
        {\mx{h}}^{\prime}(t+1)=  \mx{P}_{\mx{h}}(t+1,t) {\mx{h}}^{\prime}(t)+ \mx{C}_{\mx{h}}^{-\tfrac{1}{2}}(t+1) \bs{\xi}(t+1),
    \end{align}
    where $ \mx{P}_{\mx{h}}(t+1,t)=\mx{C}_\mx{{h}}^{-\tfrac{1}{2}}(t+1)\mx{C}_{\mx{h}}(t+1,t)\mx{C}_{\mx{h}}^{-\tfrac{1}{2}}(t)$ is the correlation matrix.

Then,
the correlation between channels at times $t+n$ and $t$ will be obtained as follows:
\begin{align}
 &\mx{P}_{\mx{h}}(t+n,t)=\mathds{E}[\mx{h}^{\prime}(t+n) {\mx{h}^{\prime}}^{\rm H}(t)]=\prod_{i=t}^{t+n-1} \mx{P}_{\mx{h}}(i+1,i).   
\end{align}
The spectral radius of the above correlation matrix can also written as
\begin{align}
\lambda_{\max}(\mx{P}_{\mx{h}}(t+n,t))
{\le}    \prod_{i=t}^{t+n-1} \lambda_{\max}(\mx{P}_{\mx{h}}(i+1,i)).
\label{eq:sp}
\end{align}
The decaying behaviour is ensured when the spectral radius of the one step covariance matrices are less than one:
\begin{align}\label{eq:decaying_condition}
   | \lambda_{\max}(\mx{P}_{\mx{h}}(t+1,t))|<1 ~~~\forall t=0,1,..
\end{align}
because in this case we have 
\[ \lim_{n\to\infty}\lambda_{\max}(\mx{P}_{\mx{h}}(t+n,t)) = 0\]
according to \eqref{eq:sp}. 

\section{Proof of Proposition \ref{prop.rician}}\label{proof.rician}

At time 
$t$, the normalized centered channel can be described as follows:
\begin{align}
\mx{h}^{\prime}(t) =\tfrac{1}{\sqrt{s(t)N_r}}\sum_{i=1}^{s(t)} {\rm e}^{j (2\pi f_d^i(t) t+ \beta_i)} \mx{a}(\theta^i_{\rm AoA}), 
\end{align}
where 
\begin{align}
 \mx{a}(\theta)=[1, {\rm e}^{j\tfrac{2 \pi d \cos(\eta-\theta)}{\lambda}},..., {\rm e}^{j\tfrac{2 \pi d \cos(\eta-\theta)(N_r-1)}{\lambda}} ] ^{\rm T}  
\end{align}
is the array response vector and $\beta_i$ is the phase of the $i$-th \ac{NLoS} path. Also, the mean of the channel is deterministic and contains the \ac{LoS} component defined as follows:
\begin{align}
\overline{\mx{h}}(t)=K_f s(t) {\rm e}^{j (2\pi f_d^0(t) t +\beta_0)} \mx{a}(\theta^0_{\rm AoA}) .  
\end{align}
Here, $K_f$ denotes the K factor of Rician distribution defined as the ratio of the LoS power to the NLoS power. Now, we can proceed to obtain the coefficient matrix between times $t_1$ and $t_2$ as follows: 
\begin{align}\label{eq:rho111}
   &\mx{P}_{\mx{h}}(t_1,t_2) =\mathds{E}[\mx{h}^{\prime}(t_1) \mx{h}^{\prime}(t_2)^{\rm H}]=\nonumber\\
   &\tfrac{1}{\sqrt{s(t_1)s(t_2)} N_r} \mathds{E}\Big[\sum_{i=1}^{s(t_1)}\sum_{l=1}^{s(t_2)}
   {\rm e}^{j \left(2\pi \left(f_d^i(t_1) t_1-f_d^l(t_2) t_2\right)+ \beta_i-\beta_l\right)}\nonumber\\
   &\mx{a}(\theta_{\rm AoA}^i(t_1))\mx{a}^{\rm H}(\theta_{\rm AoA}^l(t_2))\Big]=\nonumber\\
   &\tfrac{1}{\sqrt{s(t_1)s(t_2)} N_r} \mathds{E}\Big[\sum_{i=1}^{\min(s(t_1),s(t_2))}{\rm e}^{j \left
   (2\pi \left(f_d^i(t_1) t_1-f_d^i(t_2) t_2\right)\right)}\nonumber\\
   &\mx{a}(\theta_{\rm AoA}^i(t_1))\mx{a}^{\rm H}(\theta_{\rm AoA}^i(t_2))\Big],
\end{align}
where in the second equality, we used the fact that the phases $\beta_i$s are uniformly distributed between $[-\pi, \pi]$ and are independent from each other and also from AoAs and AoDs. This leads to 
the fact that $f_d^i(t_1) t_1-f_d^l(t_2) t_2+\beta_i-\beta_l$ is also uniformly distributed between $[-\pi, \pi]$  for $i\neq l$ and consequently  $\mathds{E}[{\rm e}^{j \left(2\pi \left(f_d^i(t_1) t_1-f_d^l(t_2) t_2\right)+\beta_i-\beta_l\right)}]=0$. 
When the number of scatterers goes to infinity, then we can use central limit theorem to approximate the $(k,l)$ element of $\mx{P}_{\mx{h}}(t_1,t_2)$ in \eqref{eq:rho111} with the following:
\begin{align}
    &\rho_{k,l}(t_1,t_2) =\nonumber\\
    &\mathds{E}\Big[
   {\rm e}^{j\tfrac{2\pi}{\lambda}\big[t_1\nu(t_1)\cos(\gamma(t_1)-\theta_{\rm AoD}(t_1))-t_2\nu(t_2)\cos(\gamma(t_2)-\theta_{\rm AoD}(t_2))\big]}\Big]
   \nonumber\\
   &\mathds{E}\Big[{\rm e}^{j\tfrac{2\pi d}{\lambda}\big[k\cos(\eta-\theta_{\rm AoA}(t_1))-l\cos(\eta-\theta_{\rm AoA}(t_2))\big]}
   \Big]\nonumber\\
   &\triangleq\rho_{\rm temporal}(t_1,t_2)\rho_{\rm spatial}(t_1,t_2,k,l)
\end{align} 
where the expectation above is over $\theta_{\rm AoA}$ and $\theta_{\rm AoD}$. In the special cases where the user and environment do not change with time, then the parameters $\nu(t)=\nu, \gamma(t)=\gamma, \theta_{\rm AoA}(t)=\theta_{\rm AoA}, \theta_{\rm AoD}(t)=\theta_{\rm AoD}$ are fixed. Define $\tau\triangleq t_2-t_1$ and $\mu\triangleq k-l$. In the specific case where AoA and AoD follows von Mises distribution given by
{\small{
\begin{align}
    &f(\theta_{\rm AoD})=\tfrac{{\rm e}^{\kappa_{\rm AoD}\cos(\theta_{\rm AoD}-\theta^c_{AoD})}}{2\pi J_0(j\kappa_{AoD})}, f(\theta_{\rm AoA})=\tfrac{{\rm e}^{\kappa_{\rm AoA}\cos(\theta_{\rm AoA}-\theta^c_{AoA})}}{2\pi J_0(j\kappa_{AoA})},  
\end{align}}}
where $\theta^c_{AoD}$ and $\theta^c_{AoA}$ are the central angles showing the mean of the von Mises distribution. $\kappa_{\rm AoD}$ and $\kappa_{\rm AoA}$ are measures of concentration of the \ac{AoD} and \ac{AoA} around the central angles. By having this information, we can now calculate $\rho_{\rm temporal}(t_1,t_2)=\rho_{\rm temporal}(\tau)$ in closed-form as follows:
\begin{align}
  &\rho_{\rm temporal}(\tau) =\mathds{E}\Big[
   {\rm e}^{-j\tfrac{2\pi}{\lambda}\big[\tau\nu\cos(\gamma-\theta_{\rm AoD})\big]}\Big]=\nonumber\\
   &\int_{-\pi}^{\pi}  {\rm e}^{-j\tfrac{2\pi}{\lambda}\big[\tau\nu\cos(\gamma-\theta)\big]} f(\theta) {\rm d}\theta= \nonumber\\
   &\int_{-\pi}^{\pi} \tfrac{{\rm e}^{-j\tfrac{2\pi}{\lambda} \tau \nu\cos(\gamma-\theta)+\kappa_{\rm AoD}\cos(\theta-\theta^c_{AoD}) 
   }}{2\pi J_0(j\kappa_{AoD})}{\rm d}\theta\nonumber\\
   &=\tfrac{J_0\left(j\sqrt{\kappa_{AoD}^2-\tfrac{4\pi^2}{\lambda^2}\tau^2\nu^2-j\tfrac{4\pi}{\lambda}\tau \nu \kappa_{AoD}\cos(\gamma-\theta^c_{AoD})   }\right)
   }{J_0(j\kappa_{AoD})}
\end{align}

where we used the trigonometric property \cite{harmonic2017} and the definition of Bessel function of zero kind.
With a similar approach, we can obtain spatial correlation as follows: 
\begin{align}
    &\rho_{\rm spatial}(t_1,t_2,k,l)=\rho_{\rm spatial}(\mu)=\nonumber\\
    &\tfrac{J_0\left(j\sqrt{\kappa_{AoA}^2-\tfrac{4\pi^2}{\lambda^2}\mu^2 d^2+j\tfrac{4\pi}{\lambda} \mu d\kappa_{AoA}\cos(\eta-\theta^c_{AoA})   }\right)
   }{J_0(j\kappa_{AoA})},
\end{align}
when $\kappa_{AoD}=\kappa_{AoA}=0$, we have a uniform distribution for AoD and AoA. In this simple case, the correlation becomes 
\begin{align}
   \rho(\tau,\mu)=J_0\left(-\tfrac{2\pi}{\lambda}\tau\nu\right)J_0\left(-\tfrac{2\pi}{\lambda}d \mu\right),
\end{align}
which aligns with the result of \cite[Eq. 4]{baddour2004accurate}.

\section{Proof of Lemma \ref{lem:mmsechannel}}\label{proof.channel estimate}
\begin{proof}
To find the linear \ac{LMMSE} channel estimate, we need to solve the following optimization problem:
\begin{align}
    \min_{\mx{W},\mx{b}}\|\mx{h}(i)-\widehat{\mx{h}}(i)\|_2 ~~\text{s.t.}~ \widehat{\mx{h}}(i)=\mx{W} \mx{y}_p+\mx{b},
\end{align}
which by \cite{Kay88} leads to 
\begin{align}\label{eqn:rel0}
\widehat{\mx{h}}(i)=\mx{C}_{\mx{h}(i),\mx{y}_p} \mx{C}_{\mx{y}_p}^{-1}({\mx{y}}_{\rm p}-\mathds{E}[\mx{y}_{\rm p})]+\mathds{E}(\mx{h}(i)).
\end{align}
With the measurements ${\mx{y}}_{\rm p}$ given in \eqref{eqn:measurements_pilot_vectorized}, the covariance matrix $\mx{C}_{\mx{h}(i),\mx{y}_{\rm p}}$ can be given by:
\begin{align}\label{eqn:rel1}
&\mx{C}_{\mx{h}(i),{\mx{y}}_{\rm p}}=\alpha \sqrt{P_{\rm p}} \mathds{E}[(\mx{h}(i)-\overline{\mx{h}}(i))({\mx{h}}_p-\overline{\mx{h}}(i)\otimes \mx{1}_{3})^{\rm H}]\widetilde{\mx{S}}^{\rm H}\nonumber\\
&\alpha_1 \sqrt{P_{\rm p}} \mx{C}_{\mx{h}(i),{\mx{h}_p}}\widetilde{\mx{S}}^{\rm H}=\alpha_1 \sqrt{P_{\rm p}} \mx{E}_m(\mx{q},i)\widetilde{\mx{S}}^{\rm H},
\end{align}
where the last step comes from the definition \eqref{eqn:E_m}.
Assuming that the channel vector is independent of zero-mean noise vector and that $\mathds{E}[\mx{h}_p]=\overline{\mx{h}}(i)\otimes \mx{1}_{3}$, the autocovariance of $\mx{y}_{\rm p}$ is given by:
\begin{align}\label{eqn:rel2}
&\mx{C}_{\mx{y}_{\rm p}}=\mathds{E}[\widetilde{\mx{y}}_{\rm p}\widetilde{\mx{y}}_{\rm p}^{\rm H}]=\nonumber\\
&\alpha_1^2 P_{\rm p} \widetilde{\mx{S}}\mathds{E}[({\mx{h}_p}-\overline{\mx{h}}(i)\otimes \mx{1}_{3})({\mx{h}_p}-\overline{\mx{h}}(i)\otimes \mx{1}_{3})^{\rm H}]\widetilde{\mx{S}}^{\rm H}+\nonumber\\
&\sigma^2_{\rm p}\mx{I}_{3N_r\tau_{\rm p}}=\alpha_1^2 P_{\rm p}\widetilde{\mx{S}} \mx{M}_m(i)\widetilde{\mx{S}}^{\rm H}+\sigma^2_{\rm p}\mx{I}_{3N_r\tau_{\rm p}},
\end{align}
where the last step uses the definition \eqref{eqn:M_m}. By replacing \eqref{eqn:rel1} and \eqref{eqn:rel2} into \eqref{eqn:rel0}, it follows that:
\begin{align}\label{eqn:rel3}
 \widehat{\mx{h}}_m(i)&=\tfrac{1}{\alpha_1 \sqrt{P_{\rm p}}} \mx{E}_m(\mx{q},i)\widetilde{\mx{S}}^{\rm H} 
 \left(\widetilde{\mx{S}} \mx{M}_m(i)\widetilde{\mx{S}}^{\rm H}+\tfrac{\sigma^2_{\rm p}}{\alpha_1^2 P_{\rm p}}\mx{I}_{3N_r\tau_{\rm p}}\right)^{-1} \nonumber\\
 &\widetilde{\mx{y}}_{\rm p} +\overline{\mx{h}}(i) 
\end{align}
By using Woodbury matrix lemma \cite{horn2013matrix}, we can write:
\begin{align}\label{eqn:rel4}
   & \left(\widetilde{\mx{S}} \mx{M}_m(i)\widetilde{\mx{S}}^{\rm H}+\tfrac{\sigma^2_{\rm p}}{\alpha_1^2 P_{\rm p}}\mx{I}_{3N_r\tau_{\rm p}}\right)^{-1} =
    \tfrac{\alpha_1^2 P_{\rm p}}{\sigma^2_{\rm p}}\mx{I}_{3N_r\tau_{\rm p}}-\nonumber\\
    &\left(\tfrac{\alpha_1^2 P_{\rm p}}{\sigma^2_{\rm p}}\right)^2\mx{I}_{3N_r\tau_{\rm p}} \widetilde{\mx{S}}
    \left( \mx{M}_m^{-1}(i)+\widetilde{\mx{S}}^{\rm H}\widetilde{\mx{S}}\tfrac{\alpha_1^2 P_{\rm p}}{\sigma_{\rm p}^2}  \right)^{-1} \widetilde{\mx{S}}^{\rm H}
\end{align}
Moreover, by having $\widetilde{\mx{S}}^{\rm H}\widetilde{\mx{S}}=\tau_{\rm p}\mx{I}_{3N_r}$ and using again Woodbury matrix lemma, it follows that:
\begin{align}\label{eqn:rel5}
&\left( \mx{M}_m^{-1}(i)+\tfrac{\alpha_1^2 P_{\rm p} \tau_{\rm p}}{\sigma_{\rm p}^2} \mx{I}_{3N_r}  \right)^{-1} =\tfrac{\sigma^2_{\rm p}}{\alpha_1^2 P_{\rm p} \tau_{\rm p}}\mx{I}_{3N_r}-\nonumber\\
&
\left(\tfrac{\sigma^2_{\rm p}}{\alpha_1^2 P_{\rm p} \tau_{\rm p}}\right)^2\left(\mx{M}_m(i)+\tfrac{\sigma^2_{\rm p}}{\alpha_1^2 P_{\rm p} \tau_{\rm p}}\mx{I}_{3N_r}\right)^{-1}.
\end{align}
By replacing \eqref{eqn:rel5} into \eqref{eqn:rel4}, it holds that
\begin{align}
    &\left(\widetilde{\mx{S}} \mx{M}_m(i)\widetilde{\mx{S}}^{\rm H}+\tfrac{\sigma^2_{\rm p}}{\alpha_1^2 P_{\rm p}}\mx{I}_{3N_r\tau_{\rm p}}\right)^{-1}=\tfrac{\sigma^2_{\rm p}}{\alpha_1^2 P_{\rm p} \tau_{\rm p}}\mx{I}-\nonumber\\
    &\tfrac{\sigma^2_{\rm p}}{\alpha_1^2 P_{\rm p} \tau_{\rm p}}\widetilde{\mx{S}}\widetilde{\mx{S}}^{\rm H}+\tfrac{1}{\tau_{\rm p}^2} \widetilde{\mx{S}}\left(\mx{M}_m(i)+\tfrac{\sigma^2_{\rm p}}{\alpha_1^2 P_{\rm p} \tau_{\rm p}}\mx{I}\right)^{-1}\widetilde{\mx{S}}^{\rm H}.
\end{align}
Multiplying the latter relation by $\widetilde{\mx{S}}^{\rm H}$ leads to
\begin{align}\label{eqn:rel6}
    &\widetilde{\mx{S}}^{\rm H}\left(\widetilde{\mx{S}} \mx{M}_m(i)\widetilde{\mx{S}}^{\rm H}+\tfrac{\sigma^2_{\rm p}}{\alpha_1^2 P_{\rm p}}\mx{I}_{3N_r\tau_{\rm p}}\right)^{-1}
    =\nonumber\\
    &\tfrac{1}{\tau_{\rm p}} \left(\mx{M}_m(i)+\tfrac{\sigma^2_{\rm p}}{\alpha_1^2 P_{\rm p} \tau_{\rm p}}\mx{I}_{3N_r}\right)^{-1}\widetilde{\mx{S}}^{\rm H}.
\end{align}
This together with \eqref{eqn:rel3} gives the final result. 
\end{proof}

\section{Proof of Theorem \ref{thm.Instantanous_SINR} }\label{proof.thm.SINR}
Based on the received measurements at data time slots, the \ac{BS} employs the optimal \ac{MMSE} receiver to estimate the transmitted data vector of the tagged user in time slot $i$ of the $m$-th frame, i.e., $\widehat{{x}}(i)=\mx{g}^{\star}(\mx{q},i)\mx{y}$.
The expected power of $\widehat{{x}}(i)$ conditioned on the prior information provided in $\bs{\zeta(i)}$, is given by:
\begin{align}\label{eq:data_estimate_power}
 \mathds{E}_{\widehat{{x}},\mx{n}_{\rm d},\mx{h}(i)|\zeta(i)}  \left[|\widehat{{x}}|^2\right]=\left\langle \mx{g}^{\rm H}\mx{g}, \mathds{E}\left[\mx{y}\mx{y}^{\rm H}| \bs{\zeta}(i) \right]\right\rangle . 
\end{align}
The expression $\mathds{E}\mx{y}\mx{y}^{\rm H}| \bs{\zeta}(i)$ in the above formula can be stated as
 \begin{align}\label{eq:Eyy^H|zeta}
     \mathds{E}[\mx{y}\mx{y}^{\rm H }|\bs{\zeta}_k(i)]=\sum_{k=1}^K \alpha_k^2 P_{{\rm d}, k} \mx{D}_k+\sigma_{\rm d}^2\mx{I}_{N_r},
 \end{align}
 where
 \begin{align}\label{eq:D_k_expression}
 &\mx{D}_k\triangleq\mathds{E}\left[\mx{h}_k(i)\mx{h}_k(i)^{\rm H }| \bs{\zeta}_k(i)\right]  =\mx{C}_{\mx{h}_k(i)|\bs{\zeta}_k(i)}+\nonumber\\
 &\mathds{E}[\mx{h}_k(i)|\bs{\zeta}_k(i)]
 \mathds{E}[\mx{h}_k(i)|\bs{\zeta}_k(i)]^{\rm H }=\mx{Q}_k+\mx{z}_k\mx{z}_k^{\rm H },
 \end{align}
 and $\mx{z}_k=\overline{\mx{h}}_k(i)+\bs{\Psi}_k(\bs{\zeta}_k(i)-\overline{\mx{h}}(i)\otimes \mx{1}_{2})$.
The last equality in \eqref{eq:D_k_expression} comes from \cite[Proposition 1]{Fodor:2021}.
By replacing \eqref{eq:Eyy^H|zeta} into \eqref{eq:data_estimate_power}, it follows that:
\begin{align}\label{eq:E|x|^2}
      &  \mathds{E}_{\widehat{{x}},\mx{n}_{\rm d},\mx{h}(i)|\bs{\zeta}(i)}  \left[|\widehat{{x}}|^2\right]=\sum_{k=1}^K \alpha_k^2 P_{{\rm d}, k} \langle \mx{g}^{\rm H }\mx{g}, \mx{D}_k\rangle+\sigma_{\rm d}^2\|\mx{g}\|_2^2.
\end{align}

We can also rewrite \eqref{eq:E|x|^2} as follows:

\begin{align}
   &  \mathds{E}_{\widehat{{x}},\mx{n}_{\rm d},\mx{h}(i)|\bs{\zeta}(i)} \left[ |\widehat{{x}}|^2\right]=
   \alpha^2 {P_{d}}_1 \left\langle \mx{z}_1\mx{z}_1^{\rm H }, \mx{g}^{\rm H }\mx{g} \right\rangle+\nonumber\\
   &\sum_{k=2}^K\alpha_k^2 P_{{\rm d}, k}
   \left\langle \mx{z}_k\mx{z}_k^{\rm H }, \mx{g}^{\rm H } \mx{g} \right\rangle+\sum_{k=1}^K \alpha_k^2 P_{{\rm d}, k} \left\langle \mx{Q}_k,\mx{g}^{\rm H } \mx{g} \right\rangle+\sigma_{\rm d}^2 \|\mx{g}\|_2^2.
\end{align}
By having the expected power of estimated data symbols, we can now form the instantaneous \ac{SINR} of the data estimate in slot $i$ of frame $m$ corresponding to the tagged user as in \eqref{eq:SINR1}.

{\small
\begin{align}\label{eq:SINR1}
 & \gamma(\mx{q},i,\zeta(i))\triangleq \tfrac{\alpha^2 P \left\langle \mx{z}_1\mx{z}_1^{\rm H },  \mx{g}^{\rm H }\mx{g} \right\rangle}
  {\sum_{k=2}^K\alpha_k^2 P_{{\rm d}, k}
   \left\langle \mx{z}_k\mx{z}_k^{\rm H },  \mx{g}^{\rm H } \mx{g} \right\rangle+\sum_{k=1}^K \alpha_k^2 P_{{\rm d}, k} \left\langle \mx{Q}_k, \mx{g}^{\rm H } \mx{g}  \right\rangle+\sigma_{\rm d}^2 \|\mx{g}\|_2^2}  
\end{align}}

%
By having \eqref{eq:F1} and \eqref{eq:E|x|^2} in mind, the instantaneous \ac{SINR} can be rewritten as
\begin{align}\label{eq:SINR2}
    \gamma(\mx{q},i,\bs{\zeta}(i))=\tfrac{\left\langle \mx{g}^{\rm H } \mx{g}, \mx{F}-\mx{F}_1 \right\rangle}{\left\langle \mx{g}^{\rm H } \mx{g}, \mx{F}_1 \right\rangle}.
\end{align}
It then follows that based on \eqref{eq:G^star}, $\mx{g}^{\rm H } \mx{g}$ can be stated as:
\begin{align}\label{eq:G^HG}
 \mx{g}^{\rm H } \mx{g}=\alpha_1^2 P_{\rm d} \mx{F}^{-1} \mx{z}_1 \mx{z}_1^{\rm H } \mx{F}^{-1}.  
\end{align}

By incorporating \eqref{eq:G^HG} into \eqref{eq:SINR2}, we reach to the following equation. 
    \begin{align}\label{eq:SINR3}
&\gamma(\mx{q},i,\bs{\zeta}(i))=\tfrac{\alpha_1^2 P_1 \left\langle  \mx{F}^{-1} \mx{z}_1 \mx{z}_1^{\rm H } \mx{F}^{-1} , \mx{z}_1 \mx{z}_1^{\rm H } \right\rangle}
{\left\langle \mx{F}^{-1} \mx{z}_1 \mx{z}_1^{\rm H } \mx{F}^{-1}, \mx{F}\right\rangle-\alpha_1^2 P_1 \left\langle  \mx{F}^{-1} \mx{z}_1 \mx{z}_1^{\rm H } \mx{F}^{-1} , \mx{z}_1\mx{z}_1^{\rm H } \right\rangle}.
\end{align}
 By some algebraic manipulations, the \ac{SINR} expression in \eqref{eq:SINR3} can be written as 
\begin{align}
    &\gamma(\mx{q},i,\zeta(i))=\tfrac{\alpha_1^2 P_{\rm d} \mx{z}_1^{\rm H } \mx{F}^{-1} \mx{z}_1}{1-\alpha_1^2 P_{\rm d} \mx{z}_1^{\rm H } \mx{F}^{-1} \mx{z}_1}
\end{align}

Ultimately, by leveraging the insights from \cite[Lemma 1]{Abrardo:19}, the aforementioned expression can be more succinctly streamlined to \eqref{eq:Instantanous_SINR_simplified}.  
\section{Proof of Theorem \ref{thm.stiel}}\label{proof.thm.stiel}
Define the expectation of \ac{SE} and \ac{SINR},  by
\begin{align}
&\overline{\textup{SE}}(\mx{q},i)\triangleq \mathds{E}\left[\textup{SE}(\mx{q},i,
\bs{\zeta}(i))\right], \overline{\gamma}(\mx{q},i)\triangleq  \mathds{E}\left[\gamma(\mx{q},i,\bs{\zeta}(i))\right].
\end{align}

The aim is to prove that the following limit holds with high probability. 
\begin{align}\label{eq:approx1}
    \tfrac{1}{N_r}{\textup{SE}}(\mx{q},i,
\bs{\zeta}(i))\xrightarrow{N_r \to \infty}\tfrac{1}{N_r}{\rm SE}^{\circ}(\mx{q},i),
\end{align}
where ${\rm SE}^{\circ}(\mx{q},i)\triangleq \log(1+\gamma^\circ(\mx{q},i))$ and $\gamma^\circ(\mx{q},i)$ is an expression to be defined later.
To show \eqref{eq:approx1}, first, by taking the expectation with respect to the random channel estimates and using Jensen's inequality bounds for concave functions borrowed from \cite{jensen_gap}, we have

\begin{align}\label{eq:jensen}
 &    \tfrac{1}{N_r}\left|\overline{\textup{SE}}(\mx{q},i)-\log(1+{\gamma}^{\circ}(\mx{q},i) )\right|\le\tfrac{c^\prime}{N_r}\left[\Big|\overline{\gamma}(\mx{q},i)-{\gamma}^{\circ}(\mx{q},i) \Big|\right],
\end{align}
where $c^{\prime}$  is some constant term independent of the frame parameters. Now, it remains to prove that $\tfrac{1}{N_r}\overline{\gamma}(\mx{q},i)$ is well approximated by $\tfrac{1}{N_r}
{\gamma}^{\circ}(\mx{q},i)$ when $N_r$ tends to infinity. 
We start by this fact that the instantaneous \ac{SINR} given in Theorem \ref{thm.Instantanous_SINR} concentrates around its mean. Since the interference channel and tagged user channel are assumed to be independent, we can separate the expectations over the channels corresponding to tagged user and interference. If we fix $\mx{z}_2,\ldots, \mx{z}_K$ and only take expectation over $\mx{z}_1$, due to concentration inequality provided in \cite[Lemma B.26]{bai2010spectral}, we have
\begin{align}\label{eq:expectation_over_z1}
& \tfrac{1}{N_r}{\gamma}(\mx{q},i,\bs{\zeta}(i))\xrightarrow{N_r \to \infty}\tfrac{1}{N_r}\overline{\gamma}(\mx{q},i)=\nonumber\\
&\tfrac{1}{N_r}\mathds{E}\left[{\gamma}(\mx{q},i,\bs{\zeta}(i))\Big| \mx{z}_2,\ldots, \mx{z}_K\right]
=\tfrac{1}{N_r}\left\langle \mx{R}_{\mx{z}_1}, \mathds{E}_{\mx{z}_2,\ldots, \mx{z}_K}[ \mx{F}_1^{-1}]  \right\rangle,
\end{align}
where $\mx{R}_{\mx{z}_1}=\mathds{E}[\mx{z}_1\mx{z}_1^{\rm H }]$. 
The above result is an extended version of \cite[Lemma B.26]{bai2010spectral} to the Rician case. Define $\mx{B}=\sum_{k=2}^K \alpha_k^2 P_{{\rm d}, k} \mx{z}_k \mx{z}_k^{\rm H }+\mx{S}$. By \eqref{eq:F1}, it holds that
$\mx{F}_1=\mx{B}+\mx{S}+\rho_{\rm d} \mx{I}_{N_r}$.
Now, we prove that the expression on the right-hand side of \eqref{eq:expectation_over_z1} indeed tends to a stationary point when the number of \ac{BS} antennas tends to infinity. It then follows that
\begin{align}\label{eq:t1}
&\tfrac{1}{N_r} \left\langle \mx{R}_{\mx{z}_1}, \mx{F}_1^{-1}  \right\rangle    \xrightarrow{N_r \to \infty} 
\tfrac{1}{N_r} \left\langle \mx{R}_{\mx{z}_1}, \mathds{E}_{\mx{z}_k, k=2,\dots,K} \mx{F}_1^{-1}  \right\rangle
\nonumber\\
&\xrightarrow{N_r \to \infty} \tfrac{1}{N_r}\left\langle \mx{R}_{\mx{z}_1},\mx{D}^{-1}  \right\rangle\triangleq \tfrac{1}{N_r}{\gamma}^{\circ}(\mx{q},i), 
\end{align}
where $\mx{D}\triangleq \mx{\Gamma}+\mx{S}+\rho_{\rm d} I_{N_r}$ and $\mx{\Gamma}$ is
a matrix depending on the Stieltjes transform corresponding to the empirical distribution of $\mx{B}$ and is given by
\begin{align}
    \mx{\Gamma}=\sum_{k=2}^K \tfrac{\alpha_k^2 P_{{\rm d}, k}\mx{R}_{\mx{z}_k}}{1+m_{\mx{B}_{k}}(\rho_{\rm d})},
\end{align}
where 
\begin{align}
   m_{\mx{B}_{k}}(\rho_{\rm d}) \triangleq \langle  \mx{R}_{\mx{z}_k}, (\mx{B}_{k}+\rho_{\rm d} \mx{I}_{N_r})^{-1} \rangle,
\end{align}
and $\mx{B}_{k}\triangleq \mx{B}- \mx{z}_k \mx{z}_k^{\rm H }$.
The approximation error in \eqref{eq:t1} is given by:
\begin{align}\label{eq:t3}
   w_N=\tfrac{1}{N_r}\left\langle \mx{R}_{\mx{z}_1},(\mx{B}+\rho_{\rm d} \mx{I}_{N_r})^{-1}  \right\rangle-
   \tfrac{1}{N_r} \left\langle \mx{R}_{\mx{z}_1}, \mx{D}^{-1} \right\rangle .
\end{align}
By extending the findings of \cite[Theorem 1]{Hoydis:13} to encompass the Rician channel with a \ac{LoS} component, and borrowing results of concentration inequalities in random matrix theory \cite{bai2010spectral}, it can be shown that the magnitude of $w_N$ approaches zero as $N_r$ becomes sufficiently large. Thus, the relations \eqref{eq:t1} and \eqref{eq:expectation_over_z1} show that $\tfrac{1}{N_r}\gamma(\mx{q},i,\bs{\zeta}(i))$ tends to $\tfrac{1}{N_r}\overline{\gamma}(\mx{q},i)$ with high probability and $\tfrac{1}{N_r}\overline{\gamma}(\mx{q},i)$ is asymptotically approximated by $\tfrac{1}{N_r}{\gamma}^{\circ}(\mx{q},i)$  . This together with \eqref{eq:jensen} shows the relation \eqref{eq:approx1}.

\bibliographystyle{IEEEtran}
\bibliography{samplingV3}
\end{document}